\documentclass[conference]{IEEEtran}
\IEEEoverridecommandlockouts
\usepackage[T1]{fontenc}
\usepackage[utf8]{inputenc}
\usepackage{inputenc}
\usepackage[english]{babel}

\usepackage[table]{xcolor}
\definecolor{maroon}{cmyk}{0,0.87,0.68,0.32}

\usepackage{hyperref} 
\usepackage{multirow}

\usepackage[ruled,linesnumbered]{algorithm2e}
\usepackage{amsmath,amssymb,amsfonts,amsthm}
\usepackage{tikz}
\usepackage{csquotes}
\usepackage{dashbox}
\usepackage{todonotes}
\usepackage{enumitem}   
\usepackage{xcolor}

\usepackage{graphicx,subfigure}
\definecolor{bblue}{HTML}{4F81BD}
\definecolor{rred}{HTML}{C0504D}
\definecolor{ggreen}{HTML}{9BBB59}
\definecolor{ppurple}{HTML}{9F4C7C}
\usepackage{xcolor}
\usepackage{pgfplots}

\definecolor{findOptimalPartition}{HTML}{D7191C}
\definecolor{storeClusterComponent}{HTML}{FDAE61}
\definecolor{dbscan}{HTML}{ABDDA4}
\definecolor{constructCluster}{HTML}{2B83BA}

\usepackage{trace}
\usepackage{makeidx}
\usepackage{mdframed}

\usepackage{listings}
\usepackage{color}
\usepackage{footmisc} 
\newcommand{\comment}[1]{}

\usepackage{fancyhdr} 
\usepackage{graphicx}

\usepackage{indentfirst}

\usepackage{booktabs}    

\usepackage{dashbox}
\usepackage{tikz}

\usepackage{ulem}

\usepackage[
	n,
	operators,
	advantage,
	sets,
	adversary,
	landau,
	probability,
	notions,	
	logic,
	ff,
	mm,
	primitives,
	events,
	complexity,
	asymptotics,
	keys]{cryptocode}
	
\usepackage{dashbox}
\usepackage{amsfonts}
\usepackage{booktabs}
\usepackage{siunitx}

\newdimen\R 
\newdimen\L 
\usepackage{graphicx}
\usepackage{indentfirst} 
\usepackage{framed}

\usepackage{etoolbox}
\usepackage{tkz-euclide}
\tikzset{%
    pics/sema/.style args={#1/#2/#3}{code={%
        \ifstrequal{#2}{0}{%
            \node[circle,minimum width=0.55mm,draw,fill=#1] {};
        }{%
            \tkzDefPoint(0,0){O}
            \tkzDrawSector[R,fill=#1](O,0.55mm)(90,90-#2)
            \tkzDrawSector[R,fill=#3](O,0.55mm)(90-#2,90-360)
    }
    }},
}
\usepackage{float}
\usepackage{tabularx}
\usepackage{lipsum} 

\usepackage{babel}
\usepackage[font=small,labelfont=bf,tableposition=top]{caption}
\usepackage{booktabs}
\usepackage{threeparttable}

\usepackage[most]{tcolorbox}
\usepackage{xcolor}
\usepackage{graphicx}
\usepackage{colortbl}
\usepackage{array}
\usepackage{pifont}
\usepackage{verbatim}

\usepackage{pifont}
\usepackage{xcolor}
\newcommand{\cmark}{\textcolor{green!80!black}{\ding{51}}}
\newcommand{\xmark}{\textcolor{red}{\ding{55}}}
\renewcommand{\arraystretch}{1.2}

\renewcommand*{\arraystretch}{1.5}%
\definecolor{tabred}{RGB}{230,36,0}%
\definecolor{tabgreen}{RGB}{0,116,21}%
\definecolor{taborange}{RGB}{250,124,30}%
\definecolor{tabbrown}{RGB}{171,70,0}%
\definecolor{tabyellow}{RGB}{251,253,169}%
\newcommand*{\redtriangle}{\textcolor{tabred}{\ding{115}}}%
\newcommand*{\greenbullet}{\textcolor{tabgreen}{\ding{108}}}%
\newcommand*{\orangecirc}{\textcolor{taborange}{\ding{109}}}%
\newcommand*{\headformat}[1]{{\small#1}}%
\newcommand*{\vcorr}{%
  \vadjust{\vspace{-\dp\csname @arstrutbox\endcsname}}%
  \global\let\vcorr\relax
}%

\begin{document}

\title{SoK: TEE-assisted Confidential Smart Contract \thanks{$\star$: This paper has been accepted by PETs 2022.}}

\author{\IEEEauthorblockN{Rujia Li \IEEEauthorrefmark{1}\IEEEauthorrefmark{2}\thanks{ $\eth$: These authors contributed equally to this work.}$^\eth$
Qin Wang \IEEEauthorrefmark{3}$^\eth$,
Qi Wang \IEEEauthorrefmark{1}, 
David Galindo \IEEEauthorrefmark{2},
Mark Ryan \IEEEauthorrefmark{2}}
\IEEEauthorrefmark{1} \textit{Southern University of Science and Technology}, China.\\
\IEEEauthorblockA{\IEEEauthorrefmark{2} \textit{University of Birmingham},  United Kingdom.\\
\IEEEauthorrefmark{3} \textit{CSIRO Data61}, Australia. 
}
}

\IEEEoverridecommandlockouts

\maketitle


\begin{abstract} 
The blockchain-based smart contract lacks privacy since the contract state and instruction code are exposed to the public. Combining smart-contract execution with Trusted Execution Environments (TEEs) provides an efficient solution, called \textit{TEE-assisted smart contracts}, for protecting the confidentiality of contract states. However, the combination approaches are varied, and a systematic study is absent. Newly released systems may fail to draw upon the experience learned from existing protocols, such as repeating known design mistakes or applying TEE technology in insecure ways. In this paper, we first investigate and categorize the existing systems into two types: the \textit{layer-one} solution and \textit{layer-two} solution. Then, we establish an analysis framework to capture their common lights, covering the desired properties (for contract services), threat models, and security considerations (for underlying systems). Based on our taxonomy, we identify their ideal functionalities, and uncover the fundamental flaws and reason for the challenges in each specification’s design. We believe that this work would provide a guide for the development of TEE-assisted smart contracts, as well as a framework to evaluate future TEE-assisted confidential contract systems.

\end{abstract}

\section{Introduction}
Smart contract was originally introduced by Szabo~\cite{szabo1996smart} and further developed by Ethereum~\cite{wood2014ethereum} in the blockchain systems. The blockchain-based smart contracts~\cite{delmolino2016step,hewa2021survey,alharby2017blockchain} adopt Turing-complete scripting languages to achieve complicated functionalities~\cite{jansen2019smart} and execute the predefined logic through state transition replication over consensus algorithms to realize final consistency. Smart contracts enable unfamiliar and distributed participants to fairly exchange without trusted third parties, and are further used to establish a uniform approach for developing decentralized applications (DApps~\cite{raval2016decentralized}). However, blockchain-based smart contract lacks \textit{confidentiality}. The state information and the instruction code are completely transparent~\cite{zou2019smart,zhang2019security,Goldfeder2018PrivateSC}. Any states with their changes are publicly accessible and all users' transaction data and contract variables are visible to external observers. Without privacy, building advanced DApps that rely on the user's sensitive data becomes a challenge~\cite{steffen2019zkay,baghery2019efficiency,Unterweger2018LessonsLF,zhang2016town}. For instance, smart contracts in Ethereum \cite{wood2014ethereum} cannot be directly applied to Vickrey auction \cite{blass2019borealis, galal2019trustee} or e-voting systems~\cite{cortier2016sok,Rathee2021OnTD}, where the bid and vote require to be hidden from the public. Moreover, DApps without privacy protection might be prohibited by European Union because they go against the General Data Protection Regulation~\cite{gdpr,voigt2017eu}. Thus, the complete transparency of smart contracts constrains their wide adoption. Recently, researchers have explored many cryptographic solutions to solve these issues, including utilizing techniques of zero-knowledge proof (ZKP) \cite{kosba2016hawk,kalodner2018arbitrum,baghery2019efficiency,bunz2018bulletproofs,bunz2020zether,chen2020pgc}, homomorphic encryption (HE) \cite{solomon2021smartfhe} and secure multiparty computation (MPC) \cite{zyskind2015enigma}. However, these approaches are merely applicable to applications requiring simple computations.

Moving complex computations into secure hardware can provide applications with privacy as well as good performance. The use of the trusted execution environments (TEEs)~\cite{lee2020keystone,ekberg2013trusted,kim2017enhancing,kaplan2016amd,brasser2019sanctuary} becomes thus a general-purpose solution for confidential smart contracts. The TEE is a new feature provided by recent commodity processors. It has the ability to provide secure environments for running contract code in isolation while guaranteeing execution integrity and state confidentiality. For instance, $\text{Intel}^\circledR$ Software Guard Extension ($\text{Intel}^\circledR$ SGX) \cite{mckeen2013innovative,zhao2016performance,cui2021dynamic,li2021offline} allows a user to create a secure area called~\textit{enclave}. Afterwards, the user unitizes the remote attestation protocol to prove to remote parties that the applications are indeed running inside an enclave. Then, the enclave establishes a secure channel to communicate with remote hosts, where the messages are encrypted. In this way, SGX runs trusted codes in an enclave and uses the CPU hardware to prevent attackers from seeing or tampering with sensitive data. Such a technique provides the high-level security for inside processes to resist attacks against outside software, even the most privileged instruction from the operating system. As a promising alternative, various smart contract platforms taking advantages of TEEs have been proposed, especially by companies working on consortium blockchain platforms, such as Alibaba CONFIDE~\cite{yan2020confidentiality}, Visa’s LucidiTEE~\cite{sinhaluciditee} and China's CHANG’AN Chain~\cite{unlockblockchain_2021,financials21}.

\begin{figure*}[!htb]
    \centering
    \includegraphics[width=1\linewidth]{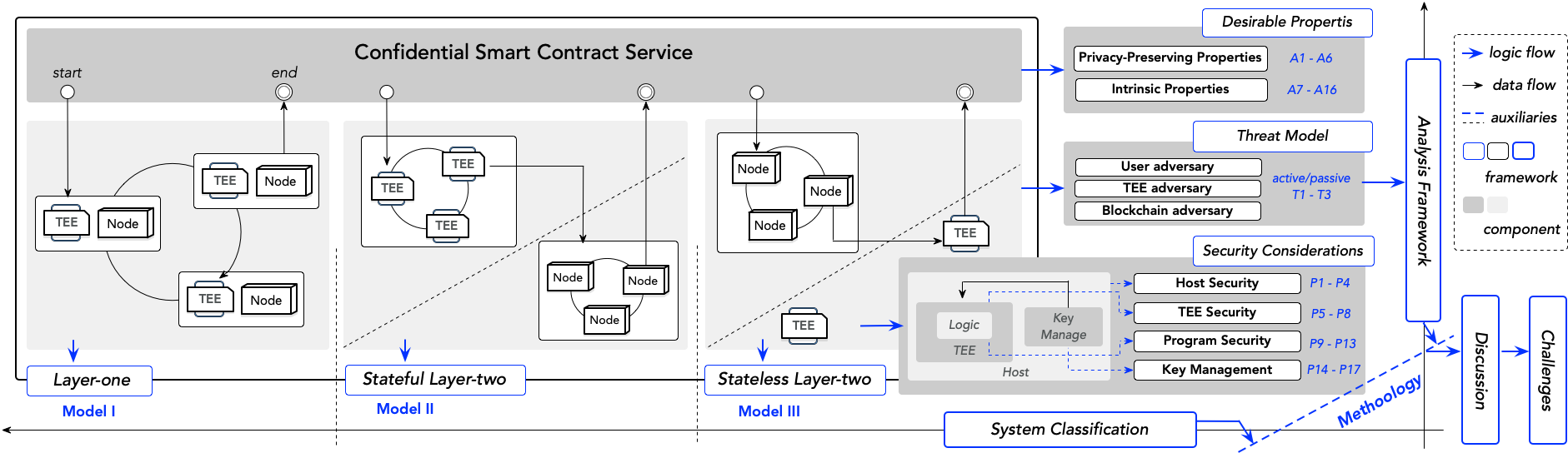}
   \caption{ \textbf{Systematization methodology.} We delineate current confidential smart contract systems along two principal axes. In the \textit{horizontal} axes, we identify two types of TEE-assisted systems according to the ways of combination. In the \textit{vertical} axes, we give our analysis framework in three aspects. The \textit{property} corresponds to the confidential smart contract service, providing the functionalities to the end-users. The \textit{threat model} and \textit{security consideration} focus on their underlying blockchain systems that support upper-layer services. With these two axes as our research methodology, we further present the related discussions and open challenges.}
    \label{fig:methodogy}
\end{figure*}

Although various TCSC protocols have been proposed, newly released projects may fail to draw upon the experience learned from existing protocols, such as repeating known design mistakes or applying cryptography in insecure ways. For example, an absence of economic incentives will pose security risks and decrease the protocol's stability. However, the recent-proposed TCSC scheme Hybridchain~\cite{wang2020hybridchain} repeats similar pitfalls by simply combining the TEE with a permissioned blockchain network, omitting considerations on the miner's incentive mechanism. The repeating of pitfalls comes from twofold. Firstly, in-the-wild projects differ from one to another, and a relatively unique model is absent, which narrows the developers' vision. Meanwhile, a unified evaluation framework is missing, causing many security threats to be uncovered and resulting in considerable loss from applications underpinning the execution of confidential smart contracts. This paper aims to abstract a high-level framework to simply and clearly systematize knowledge on current TCSC schemes. We attempt to capture some commonalities among these projects regarding their features, properties, and potential security vulnerability. We believe that establishing evaluation criteria to measure features and identify problems and flaws of existing TCSC protocols will offer a good guide for industry communities and promote the DApps prosperity. Main contributions (a visualized guideline in Fig.\ref{fig:methodogy}) are:

\begin{itemize}
    \item We provide a systematization of existing  TCSC systems driven from academic work and \textit{in production} projects. Based on their operating mechanisms and ways of combination, we investigate and categorize a set of typical protocols into two main classifications: the \textit{layer-one} solution and the \textit{layer-two} solution.
    
    \item We establish a unified evaluation framework for confidential smart contract systems. We consider two parts: the smart contracts used as \textit{service}s, and underlying supported blockchain \textit{system}s. Accordingly, the framework covers three aspects: \textit{desirable properties} for contract services, \textit{threat model} and \textit{security consideration} for underlying systems. Specifically, we discuss two different types of desirable properties: \textit{typical properties} that inherit from traditional smart contracts and featured \textit{privacy-related properties}. Then, we emphasize practical issues, pitfalls, and remedies in designing TEE-assisted blockchains from four aspects (\textit{host}/\textit{TEE}/\textit{program} securities and \textit{key management} services). 
    
    \item We conduct a comparative analysis of existing protocols based on our evaluation framework. We discuss systems both from their \textit{common designs} (system classification, threat model) and \textit{distinguishing features} (designs, properties). The common designs show us the consistent idea when re-designing the system, while the distinguished features highlight the ingenuity of each system design that deviates from others (see Tab.\ref{tab-properties}/Tab.\ref{tab-pitfall}).
    
    \item We further give a comprehensive discussion of current designs and implementations, including a running example, comparisons between layer-one and layer-two systems from the perspectives of \textit{security}, \textit{efficiency} and \textit{easy-adoption}, and common issues on \textit{public verifiability}. Unfortunately, a mature design is still not ready for large-scale applications. We thereby point out \textit{research challenges} in this field, wishing to give insights for communities on defining their models and discovering possible solutions of designing TCSC systems. 
\end{itemize}

The rest of the paper is organized as follows.
Sec.\ref{sec-tour} gives a high-level introduction on how to operate a confidential smart contract inside TEEs. Sec.\ref{sec-methodology} provides the systematization methodology (\textit{system classification} and \textit{evaluation framework}). \textit{Layer-one} and \textit{layer-two} systems are analysed in Sec.\ref{sec-layer1} and Sec.\ref{sec-layer2}, respectively. Discussions are provided in Sec.\ref{sec-discussion}. Research challenges are summarised in Sec.\ref{sec-chall}. Finally, Sec.\ref{sec-conclu} gives concluding remarks. Supplementary details are stated in Appendix A-D.

\section{A Lightning Tour}
\label{sec-tour}
\begin{figure}[!htb]
    \centering
    \includegraphics[width=0.95\linewidth]{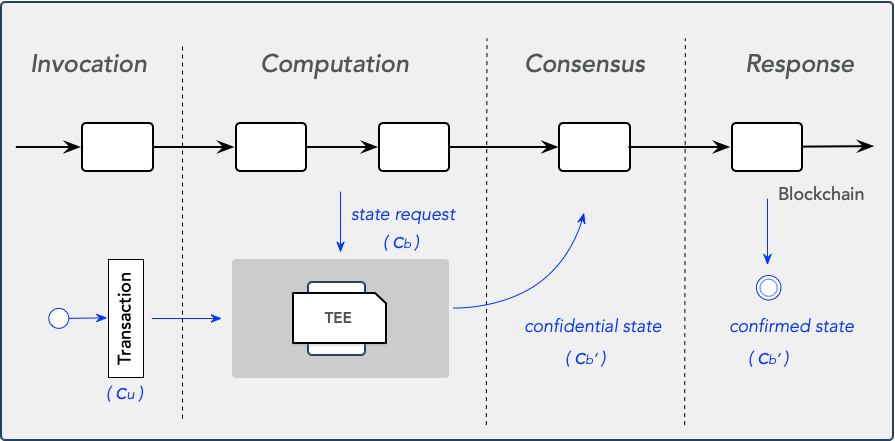}
   \caption{TEE-assisted confidential smart contract workflow.} 
    \label{fig:tour}
\end{figure}

This section gives a high-level description and offers a running example to illustrate how a typical confidential smart contract operates.
From existing literature~\cite{cheng2019ekiden,das2019fastkitten,wang2020hybridchain,muller2020tz4fabric,russinovich2019ccf,bowman2018private}, establishing a confidential smart contract mainly requires four steps, namely \textit{invocation}, \textit{computation}, \textit{consensus} and \textit{response} (see Fig.~\ref{fig:tour}).

\subsection{Overview}

\noindent\hangindent 1em \textbf{Invocation.} In current blockchain systems, once a contract is deployed successfully, the initial state and operational code are replicated to distributed nodes. The state transition must be based on an external message call that is represented as a transaction $\mathsf{Tx}$ sent from a user. The TEE-assisted confidential contract, as a special type of smart contract, inherits the state-triggering mechanism. The major difference between confidential contracts and original protocols lies in whether a transaction has to carry the ciphertext $c_{u}$ encrypted by the TEE's public key.

\noindent\hangindent 1em \textbf{Computation.} Once receiving an invocation request ($\mathsf{Tx}$ with an encrypted argument of $c_{u}$) from the user, a TEE decrypts the ciphertext $c_{u}$ and loads the contract source code and current encrypted contract state $c_{b}$ fetched from the blockchain. Then, the TEE decrypts the state $c_{b}$ using a TEE service key, executes the contract logic, and outputs an execution result $m_{b}$. Afterwards, the TEE encrypts $m_{b}$ with a specific user's public key and obtains the ciphertext $c_{b}'$. Next, the TEE sends $c_{b}'$ to the blockchain network.

\noindent\hangindent 1em\textbf{Consensus.} 
After obtaining the encrypted state $c_{b}'$ carried by $\mathsf{Tx}$, the consensus algorithm starts to reach an agreement over distributed nodes. In particular, when a blockchain node receives a newly mined block, it will re-execute all transactions inside the block. When a majority of consensus nodes receive the same block and re-execute included transactions, the state $c_{b}'$ with its carrier $\mathsf{Tx}$ is deemed to be confirmed and becomes immutable. 

\noindent\hangindent 1em\textbf{Response.} The blockchain returns the final state $c_{b}'$ and its corresponding transaction $\mathsf{Tx}$ to the user, and this user decrypts the ciphertext $c_{b}'$ to obtain the final state. To be emphasized, even the state $c_{b}'$ is publicly accessible, only the user who owns the private key can obtain the final plaintext. 

\subsection{An Running Example}
\label{run-example}

From a bird's eye view, a TCSC can be used as an ideal contract-based \textit{black box} \cite{young1996dark} with secrecy and correctness. This idea has been adopted by several advanced security protocols~\cite{li2019auditable,li2020accountable}. We provide a secret e-voting example borrowed from Oasislabs~\cite{oasislab}.

In this example, the number of voter's choices is not allowed to be revealed until the voting is finished. Meanwhile, the voter does not want other participants to know her choice. A high-level overview is: a voter calls the contract inside TEEs by sending a transaction with an encrypted argument $c_{u}$. Next, the TEE decrypts the argument $c_{u}$ and decrypts the current encrypted state $c_{b}$ using the service key (see Tab.\ref{tab-keys} and Fig.~\ref{fig:key-use} in Appendix A). Afterward, TEE confidentially executes the voting logic and correspondingly returns the message $m_{b}'$. Then, the TEE encrypts $m_{b}'$ as $c_{b}'$, and sends $c_{b}'$ to the blockchain. Eventually, the voter fetches the final result $c_{b}'$ from the blockchain and decrypts it with her private key to obtain the voting result $voteresult$ (see Tab.\ref{tab-voting}).

\begin{table}[hpt!]
\centering
\caption{Data workflow of an e-voting protocol based on TEE-assisted confidential smart contracts.}\label{tab-voting}
\resizebox{0.95\linewidth}{!}{ 
\begin{tabular}{rp{0.25\linewidth}p{0.25\linewidth}p{0.18\linewidth}}
\toprule 
\multicolumn{1}{c}{\textbf{Stage}} &\multicolumn{1}{c}{\textbf{Voter}}  & \multicolumn{1}{c}{\textbf{TEE}} & \textbf{Blockchain} \\
\midrule
\textit{Invocation} &  $votedata \to c_{u}$;
$c_{u} \to \mathsf{Tx}$ &  & $c_{b}$; \\
\textit{Computation} &  & \cellcolor{black!20} $c_{u} \to data$; $c_{b} \to m_{b}$; $data,m_{b} \to m_{b}'$; $m_{b}' \to c_{b}'$; &  \\
\textit{Consensus} & &  & $\mathsf{Tx} \to \mathsf{B}$; $c_{b}' \to \mathsf{B}$;  \\
\textit{Response} & $c_{b}' \to voteresult$ &  &  \\
\bottomrule
\end{tabular}
}
\end{table}

A TCSC can be well qualified for the role of \textit{decentralized vote manager} in an e-voting system~\cite{cortier2016sok,cortier2014election}. Once a contract-based manager is deployed successfully, the voting logic is loaded into a TEE and corresponding secret keys are privately generated and stored inside TEEs. The encrypted state is then confirmed by the blockchain nodes. This offers the e-voting protocol with \textit{confidentiality}, \textit{neutrality}, \textit{auditability} and \textit{accountability}. Firstly, the voter's input $c_{u}$ is encrypted, and intermediate parameters (e.g., $m_{b}$) are privately processed through TEEs. External attackers cannot obtain the knowledge of sensitive information, and thus the confidentiality is achieved. Secondly, the predefined voting logic only occurs in the decentralized network when certain conditions are satisfied, bringing neutrality for the access control management. Thirdly, if a voter wants to vote for a candidate, she needs to in advance build a channel to the TEE and then send a transaction $\mathsf{Tx}$ to call the contract. Due to the protection of encrypted channels, transaction arguments are kept secret. Meanwhile, such invoking records in the form of transactions remain visible and will become immutable, ensuring the voting process accountable. Unfortunately, \textit{verifiability}, as one of fundamental properties, performs not smooth in the context of encryption. Contracts that are executed inside TEEs make the execution procedures lack public verifiability. Only the nodes who install TEEs with correct corresponding keys can verify the correctness of contract executions. However, the metadata of the transaction $\mathsf{Tx}$ retains unencrypted, making it possible to verify the absence of double spending.

\section{Systematization Methodology}
\label{sec-methodology}

To find common aspects (e.g., offered functionality, design model, adversary model), we extract recurring design patterns from publicly available publications and projects, focusing on systematization and evaluation of desirable properties (the main target of TCSC) and potential pitfalls of underlying systems. Our systematization methodology follows the idea in~\cite{Unger2015SoKSM}: \textit{classification} and \textit{evaluation}. We firstly make a classification for the current systems and then define a framework to evaluate them. Details are presented as below.

\subsection{System Classification}
\label{subsec-classification}

We classify the existing systems into two main categories: \textit{layer-one solution} ($L_1$) and \textit{layer-two solution} ($L_2$). The layer-one solution executes the contract inside a TEE in the blockchain, requiring every blockchain node to equip a TEE. Instead, the layer-two solution decouples contract computations from the blockchain. It performs most of the smart contract computations off-chain. For a clear understanding, we make a comparison of the original blockchain (e.g., Ethereum), $L_1$ solution, $L_2$ solution. As in Tab.\ref{tab-compr}, Ethereum runs smart contracts (in EVM) and consensus procedures in the same machine of distributed nodes. All the contract and transaction operations are publicly verifiable due to their total transparency. The layer-one solution performs such operations (contract execution and consensus) in the same machine, but contract operations are separate from consensus procedures. In contrast, the layer-two solution makes both of them operate independently. Contracts are executed outside the blockchain network, while the consensus happens inside each node.

\begin{table}[!hbpt]
\centering
\newcommand{\tabincell}[2]{\begin{tabular}{@{}#1@{}}#2\end{tabular}}
\caption{A comparison of Ethereum, $L_1$ and $L_2$ solution}
\label{tab-compr}
\resizebox{0.98\linewidth}{!}{ 
\begin{tabular}[t]{l|ccc}
\toprule
\multicolumn{1}{c}{}&   \rotatebox{45}{\textbf{Ethereum}}  
&   \rotatebox{45}{\textbf{$L_1$ Solution}}  
&   \rotatebox{45}{\textbf{$L_2$ Solution}}  \\
\midrule 
\textit{EVM and consensus in same machine} & \cmark & \cmark & \xmark  \\
\textit{EVM and consensus in same TEE}  &  - & \xmark   & \xmark \\
\textit{Contract execution publicly verifiable}   & \cmark & \xmark   & \xmark  \\
\textit{Contract execution peer verifiable}  & \cmark & \cmark   & \xmark  \\
\textit{Consensus procedure publicly verifiable} & \cmark & \cmark   & \cmark  \\
\bottomrule
\end{tabular}
}
\end{table}%

\subsection{Desirable Property}

Ideally, moving smart contract executions into TEEs brings additional privacy as well as maintaining the original benefits of blockchain systems. Therefore, we have identified the desirable properties in two main categories: \textit{privacy-preserving property} and \textit{blockchain intrinsic feature}.

\smallskip
\noindent\textbf{Privacy-preserving property.} The property of confidentiality is the most distinguished feature in TCSC.

\smallskip
\noindent\hangindent 2em \textit{A1. Specification hidden.} The source code of a smart contract is hidden during the deployment and the subsequent synchronization and execution.

\noindent\hangindent 2em \textit{A2. Input privacy.} The inputs fed into a confidential smart contract are hidden from the public.

\noindent\hangindent 2em \textit{A3. Output privacy.} The outputs returned from a confidential smart contract should be kept private.

\noindent\hangindent 2em \textit{A4. Procedure privacy.} The execution procedure is hidden from unauthorized parties. An adversary cannot learn the operation knowledge inside a TEE.

\noindent\hangindent 2em \textit{A5. Address unlinkability.} The address pseudonymity does not entail strong privacy guarantees \cite{androulaki2013evaluating,meiklejohn2013fistful}.  This property prevents an adversary to learn the address linkability by observing users' activities.

\noindent\hangindent 2em \textit{A6. Address anonymity.} The contract caller's identity (a user who invokes a smart contract) is hidden from an anonymity set~\cite{bunz2020zether} (see Appendix B).

\medskip
\noindent\textbf{Blockchain intrinsic feature.} TEE-assisted smart contracts inherit features given by original blockchain systems. We summarize these features as follows.

\smallskip
\noindent\hangindent 2em \textit{A7. Code immutability.} Once a contract is successfully deployed, its source code cannot be altered.

\noindent\hangindent 2em \textit{A8. (Confidential) state consistency.} Executions happening at a certain blockchain height will output the same result across different nodes.

\noindent\hangindent 2em \textit{A9. Contract interoperability.} A smart contract can call another contract and be called by others. 

\noindent\hangindent 2em \textit{A10. High availability.} A smart contract is continuously accessible without the single point of failure.

\noindent\hangindent 2em \textit{A11. Decentralized execution.} A smart contract runs over the decentralized network.

\noindent\hangindent 2em \textit{A12. Automatic execution.} A smart contract can be automatically executed once conditions are satisfied.

\noindent\hangindent 2em \textit{A13. Gas mechanism.} Operations running on the smart contract will be charged with gas fees \cite{wood2014ethereum}.

\noindent\hangindent 2em \textit{A14. Explicit invocation.} Each invocation will be formatted as a transaction and stored on blockchain.

\noindent\hangindent 2em \textit{A15. Public verifiability.} The procedure of contract execution  and result are publicly verifiable.

\noindent\hangindent 2em \textit{A16. Consensus verifiability.} The consensus procedure on the (confidential) state is publicly verifiable.

\subsection{System Evaluation}
Essentially, all TCSC systems share the same principle: \textit{a TEE will handle the data from users. After that, encrypted data flows from the TEE to blockchain.} The TEE plays a crucial role. Thus, this part defines a framework for evaluating underlying blockchain systems from four aspects: \textit{TEE host}, \textit{TEE security}, \textit{TEE program}, and \textit{TEE key management}. This framework aims to identify potential design flaws and pitfalls based on the threat model and data workflow.

\smallskip
\noindent\textbf{Threat model.} Our threat model mainly captures three types of attackers, which are stated as follows.

\noindent\hangindent 2em \textit{T1. User adversary (active/passive).} An attacker may control network between users and TEE host nodes.

\noindent\hangindent 2em \textit{T2. TEE adversary (active/passive).} An adversary may control TEE hosts or control the network between TEE and blockchain platforms.

\noindent\hangindent 2em \textit{T3. Blockchain adversary (active/passive).} An adversary may drop, modify and delay the blockchain messages. But the majority (or two-thirds) of the blockchain nodes are assumed to be honest. 

\noindent Note that adversaries are not necessarily exclusive. In some cases, adversaries in different types may collude.

\medskip
\noindent\textbf{Security considerations.} This section defines four metrics regarding system security according to the data workflow: approaches to enhance the security of a \textit{TEE host}, countermeasures to mitigate \textit{intrinsic TEE issues}, methods to prevent \textit{program flaws or bugs} inside TEEs, and solutions to clear up the \textit{TEE key security} dilemma.

\smallskip
\noindent\underline{\textit{TEE host security.}}
A TEE and its interaction with the external environment (e.g., with users or the blockchain) are operated and controlled by a host (such as a $L_1$ blockchain node). A malicious host has abilities to abort the executions of a TEE, delay and modify inputs, or even drop any ingoing or outgoing messages. The following metrics discuss the approaches to improve the TEE host's security.

\noindent\hangindent 2em \textit{P1. Host punishment mechanism.} Penalty mechanisms to reduce the risk of doing evil by a TEE host.

\noindent\hangindent 2em \textit{P2. Host incentive mechanism.} Incentive mechanisms to promote a TEE host to behave honestly.

\noindent\hangindent 2em \textit{P3. Host fault tolerance.} Solutions to make systems continually operate despite malfunctions or failures.

\noindent\hangindent 2em \textit{P4. Host authentication.} Methods to check the identity and the capability of a TEE host.

\smallskip
\noindent\underline{\textit{TEE security.}}
A TEE has inevitable weaknesses. For example, a TEE is vulnerable to side-channel attacks~\cite{brasser2017software,xu2015controlled}. These innate weaknesses directly pose severe challenges to the design and implementation of TEE-assisted contract systems. This part defines the defence approaches against these threats.

\noindent\hangindent 2em\textit{P5. TEE attestation security.} Methods to prevent TEE attestation service from being abnormally broken.

\noindent\hangindent 2em\textit{P6. TEE memory limitation.} Methods to optimize the memory size to prevent confidential data overflow.

\noindent\hangindent 2em\textit{P7. TEE physical attacks.} Approaches to prevent physical attacks, such as the Spectre vulnerability or the Meltdown vulnerability \cite{hill2019spectre}. 

\noindent\hangindent 2em\textit{P8. TEE trusted timer.} Approaches to provide a trusted timer when running a TEE. 

\smallskip
\noindent\underline{\textit{TEE program security.}}
Even a TEE is secure as assumed, a program bug may destroy the contract's confidentiality in the real world. This part focuses on the measurements to prevent TEE programs from flaws or bugs. 

\noindent\hangindent 2em \textit{P9. Workload measurement.} The workload measurement approach to prevent an infinite loop attack.

\noindent\hangindent 2em \textit{P10. Flaws detection.} Formal techniques used for the modelling and verification of the source code of smart contracts to reduce the vulnerabilities.

\noindent\hangindent 2em \textit{P11. User query restriction.} The restriction on users' queries, aiming to avoid data leakage resulting from differential-privacy analysis \cite{dwork2008differential}.

\noindent\hangindent 2em \textit{P12. Blockchain data confirmation.} Methods for a TEE to check whether input data from blockchain has been confirmed to prevent the rollback attack~\cite{homoliak2020aquareum}. 

\noindent\hangindent 2em \textit{P13. TEE output conflicts.} Methods to avoid multiple TEEs to produce a conflict result.

\smallskip
\noindent\underline{\textit{TEE key security.}}
Various keys (cf. Appendix A) are involved in the contract execution, including TEE internal keys such as the attestation key and TEE service keys for state encryption/decryption. Since service keys directly affect the protection of contract states, the key security evaluation in this SoK mainly focuses on the generation, exchange, and storage of the TEE service key. 

\noindent\hangindent 2em \textit{P14. Distributed key protocol.} The keys of confidential contracts are managed by a distributed protocol.

\noindent\hangindent 2em \textit{P15. Key rotation protocol.} The TEE replaces an old key with a fresh key for future contract encryption.

\noindent\hangindent 2em \textit{P16. Independent contract key.} Each contract is associated with a unique key, independent from the TEE.

\noindent\hangindent 2em \textit{P17. Independent TEE key.} Each TEE has a unique key, and different contracts share the same key.

\smallskip
\noindent\textbf{Systematization summary.}  The \textit{system classification} shows a general view of the TCSC systems. \textit{Desirable property} focuses on evaluating contract service provided by a TEE-assisted blockchain system. \textit{Threat model} describes the potential threats and system assumptions. \textit{Security considerations} show the evaluating indicator for current TEE-assisted systems. In the following section~\ref{ose} and~\ref{tse}, we attempt to answer the following questions: (i) What are the potential pitfalls in each security aspect; (ii) Do these pitfalls have significant security impacts; (iii) Do the designers/developers consider these pitfalls and accordingly come up with feasible remedies in their systems; (iv) What are the remedies and do they address above problems. Note that hundreds of TCSC systems have been proposed in both industry and academia. An exhaustive analysis is undesirable and infeasible. We only selected the projects that provide publicly accessible technical reports or academic papers.

\section{Layer-One Solution}
\label{sec-layer1}
\begin{figure}[!htb]
    \centering
    \includegraphics[width=0.9\linewidth]{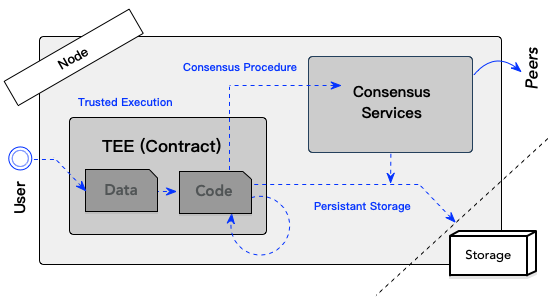}
   \caption{Layer-one execution model}
    \label{fig:on-line}
\end{figure}

The layer-one approach enables blockchain nodes to run contracts in their isolated areas, as well as conducting the consensus (see Fig.~\ref{fig:on-line}). This approach combines the consensus procedure and state execution, either in terms of logically or physically. The reason why we call this method \textit{layer-one} is that all executions are completed in the same layer of the blockchain network. The key to such an approach is to equip every blockchain node with a TEE. Indeed, this requires more integration efforts, but also comes with several advantages. The smart contract can implement stateful functionalities that receive arguments and update states instantly. In particular, a smart contract can directly access the ledger data stored in a local disk, greatly saving time often wasted in the interactive network communications.

\noindent\textbf{System model.}
In a layer-one execution model, the operation of ledger update (consensus) and state transition (contract execution) are coupled. Like Ethereum~\cite{wood2014ethereum}, smart contracts run inside blockchain nodes. Assume that a user plans to use the private contract; she only needs to upload data to the blockchain service and wait for results. The remaining procedures are completed by TEE-assisted distributed nodes. A TEE in these nodes acts as a \textit{black box} for data processing and output targeted results without the data leakage. This approach greatly improves convenience for users due to its easy access and management. As illustrated in Fig.\ref{fig:on-line}, a generic data flow goes as follows: A contract creator deploys the code into blockchain. Then, a user sends the transaction with an encrypted argument to an arbitrary blockchain node. Her request is confidentially executed inside TEEs in this node and output encrypted state. Then, the consensus algorithm in this node broadcasts the encrypted results to peers. After the encrypted results are confirmed by other blockchain nodes, users fetch on-chain results and decrypt them for the plaintext.

\subsection{Property Evaluation}

\noindent\textbf{Privacy-preserving property.} This property indicates that contract states and the procedure of contract executions are hidden from the public. To achieve privacy, layer-one systems execute these confidential contracts inside TEEs in every distributed node. CCF \cite{russinovich2019ccf}, Fabric \cite{brandenburger2018blockchain} and CONFIDE~\cite{yan2020confidentiality} follow this straightforward design where confidential contracts are loaded to the TEE of each consensus node, which encrypts both the inputs and outputs of contract states, together with their operating logic and predefined rules. Enigma\footnote{Enigma’s secret network consists of a list of secret nodes equipped with TEE, which is categorised as a layer-one solution in the context of our definition (Sec.\ref{subsec-classification}). We also note that such a secret network can be regarded as a layer-two solution in the traditional classifications in terms of Ethereum, namely, either on-Ethereum chain ($L_1$) or off-Ethereum chain ($L_2$).}~\cite{enigma} introduces the secret network and allows users to submit their transactions together with encrypted data to miners. We also notice that current layer-one solutions only focus on internal procedures rather than the linkability and anonymity of addresses and transactions. This indicates that confidential smart contracts only protect the contents that have been loaded into TEEs, while the data that relates to external users is out of the scope of this work.

\smallskip
\noindent\textbf{Blockchain intrinsic feature.} The layer-one systems inherit most of the features empowered by blockchain. More precisely, the properties of \textit{code immutability}, \textit{high availability}, \textit{explicit invocation}, \textit{decentralized execution}, \textit{automatic execution} and \textit{consensus verifiability} remain the same because basic contract executions still rely on their underlying blockchain systems. Also, the property of (confidential) \textit{state consistency} in Enigma~\cite{enigma}, CCF~\cite{russinovich2019ccf} and Fabric~\cite{brandenburger2018blockchain} remains unchanged. The states and executions from these systems follow the procedures of online consensus processes. Then, the returned results from inside TEEs still require to be confirmed on-chain. This makes their actions effectively perform the same functions as a normal smart contract, except for that the contents of states are transmitted from plaintext to ciphertext. In contrast, the property of \textit{contract interoperability} is lost since the contracts are executed in isolated TEEs. This isolation requires additional communications such as dispatching keys through the remote attestation service, bringing much complexity.

\subsection{System Evaluation}
\label{ose}

The layer-one solution encapsulates TEE computations into blockchain nodes. Every node in the network has to take responsibility for conducting confidential executions and performing the consensus. The design to coordinate TEEs and consensus within the same physical space brings many distinguished features. We start the analysis from their threat model and then dive into each component of these systems.

\smallskip
\noindent\textbf{Threat model.}  
Users in the layer-one approach are assumed to be unreliable. They may have mistakes unconsciously, like dropping messages or mis-sending transactions. Even worse, a malicious user can arbitrarily behave like faking messages, identities, or compromising other nodes. As for TEE hosts, an external attacker can monitor, eavesdrop or even compromise part of involved TEE hosts among these distributed nodes, but cannot block all traffic transmitted in communication channels. Subsequently, a TEE is supposed to work in a good condition: The attestation service is trusted, and the cryptographic primitives used inside TEEs are secure. Meanwhile, as for the blockchain network, the basic systems (ledgers) are assumed to be robust~\cite{garay2015bitcoin,garay2017bitcoin,pass2017analysis}. When running the consensus, the majority (might be two-third, depends on specific consensus algorithms) of nodes are assumed to be honest~\cite{garay2020sok}. Also, forging smart contract codes or states will happen in honest blockchain nodes with a negligible possibility. Based on that, we analyse securities from four aspects.

\smallskip
\noindent\underline{\textit{TEE host security.}} Firstly, we focus on the security of TEE hosts, or equally, individual nodes that run TEEs. Unlike classical blockchain systems, there are no explicit incentive or punishment mechanisms in this solution. This is easy to understand: A node with malicious behaviors will be instantly moved out of the committee and replaced by a new honest participant. Meanwhile, due to the fact that CCF~\cite{russinovich2019ccf} and Enigma~\cite{enigma} rely on Tendermint (a BFT variant) consensus algorithm, they can tolerate at most one-third of TEE Byzantine nodes. But the sacrifice is the increased difficulty in synchronization, especially when every node has to establish a secure channel for communications of distributed TEEs. In layer-one systems, host authentication is necessary. The node who wants to join the committee has to obtain permission from communities by proving her TEE capability. For instance, CONFIDE~\cite{yan2020confidentiality} builds a mutual authenticated protocol (MAP) (supported by SGX remote attestation techniques \cite{SGX2020}) among blockchain nodes. Any nodes joining in the network have to pass the authentication via MAP. 

\smallskip
\noindent\underline{\textit{TEE security.}} Then, we analyse TEE-level securities. Attestation service is an essential part of TEE techniques. Systems in the layer-one solution still require such services for network connection and verification. Enigma~\cite{enigma}, Fabric \cite{brandenburger2018blockchain} and CCF~\cite{russinovich2019ccf} follow the original attestation mechanism with an implicit rule: The Intel Attestation Service (IAS) should be reliable. However, this cannot be guaranteed in the case of IAS being comprised. In contrast, CONFIDE~\cite{yan2020confidentiality} utilizes a customized Decentralized Attestation Service to provide the robust authentication. As for memory limitations, layer-one systems load contract executions and consensus algorithms into one TEE-embedded node, causing an increase in disk and memory usage of individual nodes. Once the usage of TEE memory runs over the predefined settings, a decrease in the performance is inevitable~\cite{zhao2016performance}. This may further cause an unpredictably severe result like system crash-down. Fortunately, Fabric~\cite{brandenburger2018blockchain} mitigates such issues by separating the operations into two types (execution and ordering) and delays the transaction-\textit{ordering} procedures after state-\textit{execution}. Among them, only the state-\textit{execution} parts are processed inside TEEs. This decreases computation complexity and limits the memory usage to a suitable range. Physical attacks like the Spectre and Meltdown vulnerabilities \cite{hill2019spectre} are intrinsic design pitfalls that may occur inside the TEE kernel. To our knowledge, no layer-one solutions mention them or provide the remedies.

\smallskip
\noindent\underline{\textit{TEE program security.}} Next, we focus on the program-level security. Issues like overburdening may frequently happen, especially when a malicious developer deploys a contract with infinite loop logic. Unlike using the gas mechanism in Ethereum \cite{wood2014ethereum}, systems in the layer-one model constrain their running programs by the \textit{time-out} mechanism. It sets a threshold, namely, a suitable range of time that allows processing contract operations. When exceeding the time-bound, the system will abort under-processing states and restart a new round. As for the flaw detection, no formal techniques or verification tools, based on our observation, have been applied to layer-one systems. This gap needs further exploration. Similar to the previous discussion, the properties of data verification (covering both user data authenticity and blockchain data confirmation) and output conflicts are guaranteed by their underlying consensus algorithms. Each time performing the consensus, these properties are automatically checked. For instance, Enigma~\cite{enigma} relies on trusted validators, who equip with TEEs to conduct the verification procedure. Such validators maintain both the privacy of executions inside TEEs and the consistency of states that connects to peers. Once conflicts occur, validators will quickly make decisions on a block and remove another conflicting block. Fabric~\cite{brandenburger2018blockchain} performs such a process inside TEEs among committee nodes and then submits the passed results to its abstract ordering service. This service prevents forks caused by conflicting states, as well as proving a fact that: All executed messages are valid and integral once reaching the consensus agreement. It should be noted that, successful consensus procedures can merely guarantee the integrity of transactions and states, rather than linkability and authenticity that relates to physical entities.

\smallskip
\noindent\underline{\textit{TEE key management.}} Lastly, we move to the aspect of TEE key management. In layer-one systems, the key management service takes over the task of creating and managing keys for activities like attestation, verification, encryption, etc. To achieve the key management service among distributed nodes, several types of designs have been proposed. CCF~\cite{russinovich2019ccf} relies on the public key infrastructure (PKI) for certificate issuance, management, and revocation. It creates key pairs and dispatches them to every participated TEE, where each TEE holder is authenticated by the certificate. Similarly, Fabric~\cite{brandenburger2018blockchain} 
adopts an admin peer to provision the specific decryption key to $\textit{chaincode enclave}$ during bootstrapping. Enigma~\cite{enigma} setups an independent key management component to reply to the requests for encryption. Such designs help to simplify complex management procedures, as well as providing distinguishable keys for each TEE. However, these independent key management services lead to centralization even they are maintained by a group of nodes in the committee. CONFIDE~\cite{yan2020confidentiality} mitigates this issue by proposing a decentralized key management protocol. Two types of keys are involved in this protocol: the \textit{asymmetric private key} used to decrypt confidential transactions from clients and the \textit{symmetric states root key} used for state encryption/decryption between the confidential engine and storage service. 

\subsection{Pros and Cons} 
The layer-one solution provides a highly integrated approach towards confidential smart contracts.

\begin{center}
\begin{tcolorbox}[colback=gray!10,
                  colframe=black,
                  width=8cm,
                  arc=1mm, auto outer arc,
                  boxrule=0.5pt,
                 ]

This method ($L_1$) retains most blockchain features such as high availability, rollback attack resistance, decentralized execution, since the contract workflow, data structure, and usage model are consistent with existing systems.
\end{tcolorbox}
\end{center}

The layer-one solution provides a consistent interface for users without changing the customer's habits transformed from non-TEE blockchain systems. A user can use the layer-one system by directly interacting with the blockchain interface, without considering cumbersome and complicated operations between the TEE and blockchain. However, the layer-one solution still confronts several common disadvantages.

Minimizing the size of Trusted Computing Base (TCB) contributes to the TEE security~\cite{krahn2020teemon}. In particular, a small TCB has fewer errors and can reduce attack surfaces. However, complicated interactive operations for contract execution and consensus agreement in the $L_1$ solution greatly increase the size of TCB. Meanwhile, TEE products have limited secure memory. 
For example, in the current implementation of Intel SGX~\cite{cui2021dynamic}, the enclave page caches are constrained to 128 MB, and only 93 MB of those is available for applications, which limits the concurrent execution. 

Furthermore, the layer-one solution lacks compatibility, which means being incompatible with existing blockchain systems. The solution integrates the consensus procedure and the contract execution into the same blockchain node, requiring every node having to equip a TEE hardware. Nevertheless, this requirement is difficult to be fulfilled in a public blockchain while already in use (e.g., Ethereum \cite{wood2014ethereum}).

\section{Layer-Two Solution}
\label{sec-layer2}
The layer-two solution is a straightforward approach that combines the TEE and blockchain to provide smart contracts with confidentiality while keeping scalability. In such systems, the operations of smart contracts are decoupled from their underlying blockchain systems. The smart contracts are executed in an independent layer outside blockchain systems.

\begin{figure}[htb!]
    \centering
    \includegraphics[width=0.9\linewidth]{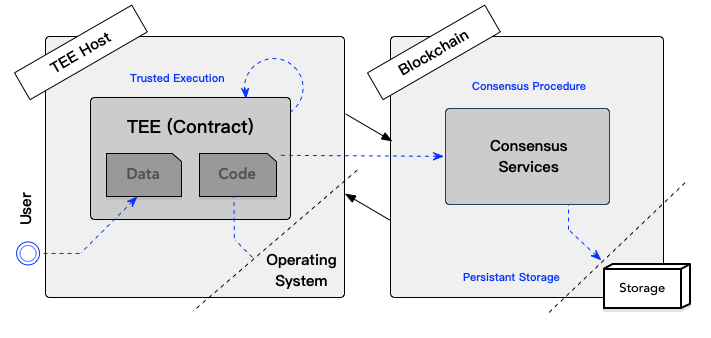}
   \caption{Layer-two execution model} 
    \label{fig:off-line}
\end{figure}

\noindent\textbf{System model.} In a general layer-two solution, the blockchain is used as a dispute resolution layer. The smart contract is executed outside the blockchain, making TEEs act as an agent between users and blockchain systems. Suppose that a user aims to use a private contract. She first needs to compile the original contract code, push binary codes to a TEE, and then upload execution results to the public ledger. As illustrated in Fig.\ref{fig:off-line}, we extract a generic data flow as follows. A user sends the encrypted input data to a TEE-powered node. Then, the TEE decrypts the input data and executes the contract. After that, the encrypted execution results are sent to the blockchain platform for verification and storage. Finally, the user fetches and decrypts the blockchain-confirmed results.

\subsection{Property Evaluation}

\noindent\textbf{Privacy-preserving property.} The \textit{confidential execution} is an essential property. In layer-two systems, such as~\cite{bowman2018private,yuan2018shadoweth, Phala2019}, the contract computations run inside Intel SGX enclaves, while TZ4Fabric~\cite{muller2020tz4fabric} moves contract executions into ARM Trusted Zone. Since the contract state-transition process happens inside TEEs, any intermediate states remain invisible to the outside. Meanwhile, to achieve the full life-cycle security for a smart contract, the input sent to a TEE and the output returned from this TEE are also required to be encrypted. For example, in ShadowEth~\cite{yuan2018shadoweth}, PDOs~\cite{bowman2018private}, Phala~\cite{Phala2019} and Hybridchain~\cite{wang2020hybridchain}, the contract invocation arguments are encrypted with the TEE public key. They can only be decrypted within the enclave. Also, before transferring execution results to the blockchain (or users), the intermediate (or final) states in an enclave are encrypted. Some variants also enhance the privacy-preserving properties from other aspects. In Phala~\cite{Phala2019}, only authorized queries to the contract will be answered. The smart contract source codes in ShadowEth~\cite{yuan2018shadoweth} are hidden during the procedures of deployment and synchronization. This further reduces the possibility of data leakage in subsequent contract executions. Considering a fixed address may expose the user who has invoked the contract, PDOs~\cite{bowman2018private} also allows the user to use pseudonym addresses for submitting a transaction (including TEE outputs) to the blockchain.

\smallskip
\noindent\textbf{Blockchain intrinsic feature.} ShadowEth~\cite{yuan2018shadoweth} and Taxa~\cite{taxa2020} introduce an external distributed service to manage the contracts, achieving the properties of \textit{code immutability}, \textit{high availability} and \textit{decentralized execution}. Meanwhile, layer-two systems satisfy \textit{state consistency} for reasons that the encrypted states of contracts in different blockchain nodes will eventually get consistent when reaching a successful agreement. Intuitively, the contracts deployed in layer-two systems should retain the features given by original blockchains. However, some fundamental properties are lost when using layer-two solutions. For example, most layer-two systems lose contract interoperability since each contract is executed in different machines. Among all the evaluated systems, only Phala~\cite{Phala2019} identifies this issue and proposes a command query responsibility segregation architecture to ensure certain interoperability. Also, public verifiability is a crucial property for the blockchain since it allows each contract invocation, and contract execution to be publicly verifiable. Unfortunately, contracts are executed in TEEs so that the outputs are encrypted. To check whether the TEE has executed contracts following loaded contract specifications is a non-trivial task.

\subsection{System Evaluation}
\label{tse}

\noindent\textbf{Threat model.} An attacker may control the network between users and TEE hosts. Meanwhile, TEEs are assumed to always produce correct results, and the smart contracts inside TEEs cannot deviate from their specifications. The main difference compared with the assumption of layer-one systems is that an adversary can observe the network activities between the TEE interfaces and active blockchain nodes.

\begin{table*}[h]
\label{tab-summary}
\begin{center}
\caption{\textbf{Desired properties for current TEE-assisted confidential smart contracts}}
\label{tab-properties}
\begin{threeparttable}
\resizebox{1\linewidth}{!}{
\begin{tabular}{ll p{3mm}p{3mm}p{3mm}p{3mm}p{3mm}p{9mm}p{3mm}p{3mm}p{3mm}p{3mm}p{3mm}p{3mm}p{3mm}p{3mm}p{3mm}p{3mm}p{3mm}}

\toprule
&\textbf{Selected Examples}& \multicolumn{6}{c}{\textbf{Privacy-Preserving Properties}} & 
 \multicolumn{9}{c}{\textbf{\;\; Blockchain Intrinsic Benefits \;\;}}
 \\
\cmidrule(lr){3-8}
\cmidrule(lr){9-19}
 & & 
\rotatebox{45}{Specification hidden} &
\rotatebox{45}{Procedure privacy} & 
\rotatebox{45}{Input privacy} &
\rotatebox{45}{Output privacy} & 
\rotatebox{45}{Address unlinkability} & 
\rotatebox{45}{Address anonymity} & 
\rotatebox{45}{Code immutable} & 
\rotatebox{45}{State consistency} & 
\rotatebox{45}{Contract interoperability} &
\rotatebox{45}{High availability.} &
\rotatebox{45}{Decentralized execution.} &
\rotatebox{45}{Automatic execution} & 
\rotatebox{45}{Gas mechanism} &
\rotatebox{45}{Explicit invocation} &
\rotatebox{45}{Public verifiability} &
\rotatebox{45}{Consensus verifiability} &
\\
\midrule 
\rowcolor{tabyellow} \textbf{Layer-one} & 2017, Enigma~\cite{enigma,TheDevel7enigma} & $\circ$ &  $\bullet$  & $\bullet$  & $\bullet$ &  $\circ$ & $\circ$ & $\downarrow$   & $-$ & $\downarrow$ & $-$ &  $-$ &  $-$& $\downarrow$ & $-$ & $\downarrow$ & $-$ \\

\rowcolor{tabyellow} & 2018, Fabric \cite{brandenburger2018blockchain}  &  $\circ$  & $\bullet$  & $\bullet$  & $\bullet$  & $\circ$ & $\circ$  & $-$ & $-$ & $\downarrow$ & $-$ & $-$   & $-$& $-$ & $-$ &$\downarrow$  & $-$ \\

\rowcolor{tabyellow}  & 2019, CCF \cite{russinovich2019ccf}  &  $\circ$  & $\bullet$    &  $\bullet$ & $\bullet$ & $\circ$ & $\circ$  &  $-$ &  $-$ & $\downarrow$ & $-$  & $-$    &$-$ & $-$ & $-$ & $\downarrow$  & $-$\\

\rowcolor{tabyellow}  &  2020, CONFIDE \cite{yan2020confidentiality} &  $\circ$  & $\bullet$ &  $\bullet$ &  $\bullet$ &$\circ$  &$\circ$  & $-$  & $-$ & $\downarrow$ & $-$  & $-$     & $-$ & $\downarrow$ & $-$ & $\downarrow$ & $-$\\

\midrule 
\textbf{Layer-two}
& 2016, Hawk~\cite{kosba2016hawk}, & $\circ$ &  $\bullet$  & $\bullet$  & $\bullet$ &  $\bullet$ & $\bullet$ & $-$ & $-$ & $-$ & $-$ & $-$  & $-$ & $-$ & $-$ & $-$ & $-$\\
& 2018, PDOs~\cite{bowman2018private}  & $\bullet$ &  $\bullet$  & $\bullet$  & $\bullet$ &  $\bullet$ & $\circ$ & $\downarrow$   &$-$ & $\downarrow$ &$\downarrow$ & $-$  & $-$ & $-$ & $-$ & $\downarrow$ & $-$ \\

& 2018, ShadowEth~\cite{yuan2018shadoweth}  & $\bullet$ &  $\bullet$  & $\bullet$  & $\bullet$ &  $\tikz\pic{sema=black/180/white};$ & $\circ$ & $-$ & $-$ & $-$ & $-$ & $-$  & $-$ & $-$ & $-$ &$\downarrow$ & $-$ \\

& 2019, Phala~\cite{Phala2019} & $\circ$ &  $\bullet$  & $\bullet$  & $\bullet$ &  $\circ$ & $\circ$ &  $-$ & $-$ & $-$ &  $\downarrow$ &  $\downarrow$  & $-$ & $-$ &$-$ & $\downarrow$ & $-$ \\

  &  2019, Ekiden~\cite{cheng2019ekiden} & $\circ$ &  $\bullet$  & $\bullet$ & $\bullet$ &  $\circ$ & $\circ$ & $\downarrow$  &$-$ & $\downarrow$ & $\downarrow$ & $-$  & $-$ & $\downarrow$  & $-$ &$\downarrow$ & $-$ \\
    
  &2019, Fastkitten~\cite{das2019fastkitten}  & $\circ$ &  $\bullet$  & $\bullet$  &   $\tikz\pic{sema=black/180/white};$ &  $\circ$ & $\circ$ & $-$   &$-$ & $\downarrow$ &$\downarrow$ & $\downarrow$  & $\downarrow$& $-$& $\downarrow$ &$\downarrow$ & $-$ \\
    
  &2019, Avalon~\cite{avalon19} & $\circ$  & $\bullet$   &  $\bullet$ & $\bullet$ & $\circ$  & $\circ$ &  $-$  & $-$ & $-$ &  $\downarrow$ &  $\downarrow$  &  $\downarrow$ &  $\downarrow$ &  $-$  &  $\downarrow$ & $-$ \\
  
  &2020, Hybridchain~\cite{wang2020hybridchain} & $\circ$ &  $\bullet$  & $\bullet$  &   $\bullet$ &  $\circ$ & $\circ$ & $-$   &$-$ & $\downarrow$ &$\downarrow$ & $\downarrow$  & $\downarrow$ & $\downarrow$& $-$  & $\downarrow$ & $-$ \\
  
   &2020, COMMITEE~\cite{Erwig2020CommiTEEAE} & $\circ$ &  $\bullet$  & $\bullet$  & $\bullet$ &  $\circ$ & $\circ$ & $\downarrow$   & $-$ & $\downarrow$ & $\downarrow$ & $\downarrow$  & $\downarrow$& $\downarrow$ & $\downarrow$ &$\downarrow$ & $-$\\
   
  &2020, PrivacyGuard~\cite{xiao2020privacyguard} & $\circ$ &  $\bullet$  & $\bullet$  & $\bullet$ &  $\circ$ & $\circ$ & $\downarrow$   & $-$ & $\downarrow$ & $\downarrow$ & $\downarrow$  & $\downarrow$& $\downarrow$ & $-$  &$\downarrow$ & $-$
   \\
  &2020, TZ4Fabric~\cite{muller2020tz4fabric}  & $\circ$ &  $\bullet$  &  $\bullet$  &   $\bullet$ &  $\circ$ & $\circ$ & $-$   & $-$ & $\downarrow$ &$\downarrow$ & $\downarrow$  & $\downarrow$ & $\downarrow$ & $-$  &$\downarrow$ & $-$\\
  
  &2020, Taxa~\cite{taxa2020}  & $\circ$ &  $\bullet$  & $\bullet$  & $\bullet$ &  $\circ$ & $\circ$ & $\downarrow$ & $-$ & $-$ &$\downarrow$ & $-$  & $-$ & $\downarrow$ & $-$ &$\downarrow$ & $-$ \\

  &2021, Erdstall~\cite{Introduc42,Technolo53} & $\circ$ &  $\bullet$  & $\bullet$  & $\bullet$ &  $\circ$ & $\circ$ & $\downarrow$ & $-$ & $\downarrow$ & $\downarrow$ & $\downarrow$  & $\downarrow$& $\downarrow$ & $\downarrow$  & $\downarrow$ & $-$
   \\

\bottomrule 
\end{tabular} 
}
     \begin{tablenotes}
       \footnotesize
       \item $\bullet$ \, Support;
       \tikz\pic{sema=black/180/white}; \,  Optionally support; $\circ$ \,  Did not support.
       \item 
       $-$ \, Benefit unchanged;  $\downarrow$ \, Benefit weakened; \, Layer-one solutions are in \colorbox{tabyellow}{\color{black}Yellow} background.
     \end{tablenotes}
   \end{threeparttable}
\end{center}
\end{table*}

\smallskip
\noindent\underline{\textit{TEE host security.}} Several layer-two solutions adopt incentive or punishment mechanisms to encourage TEE hosts to provide a stable and secure environment for executing confidential contracts. For example, Fastkitten~\cite{das2019fastkitten} and Erdstall~\cite{Introduc42,Technolo53} propose \textit{penalty} transactions, in which a host will be punished if its malicious behavior has been identified. In particular, if the TEE execution is aborted, the host will be charged according to previous deposits. In Taxa~\cite{taxa2020}, every node can identify any faulty nodes with reliable proofs for executing further economic punishment. On another route, TEE hosts in Phala~\cite{Phala2019} will get paid by providing their computing resources to users. Similarly, the remuneration in ShadowEth~\cite{yuan2018shadoweth} will be transferred to TEE hosts who execute private contracts. These mechanisms can effectively prevent malicious TEE hosts from an economic aspect. However, they are powerless against external threats. An adversary may directly terminate a TEE host at any time. Even worse, the TEE provides users with an open interface that is vulnerable to DoS \cite{liu2009research} or single-point attack. To overcome such issues and achieve fault tolerance, different methods are proposed. Fastkitten provides low-level fault tolerance by periodically saving an encrypted snapshot of current states in enclaves. If the enclave fails, the TEE host can instantiate a new enclave and restart the computation starting from the encrypted snapshot. Similarly, Taxa~\cite{taxa2020} stores a session file for maintaining and recovering user's requests. However, a malicious attacker may directly terminate the TEE host, and Fastkitten does not tolerate such host failures. Another technical route is to maintain a secure network. ShadowEth maintains a group of TEE nodes to ensure consistency via a Paxos-like~\cite{de2000revisiting} algorithm. Taxa adopts TEE-enabled computing nodes powered by a PBFT-derived PoS~\cite{gavzi2019proof} algorithm. Any node in the network has the same responsibility to privately execute smart contracts and transfer execution results to the blockchain. However, this brings additional authentication issues. A TEE host must be carefully authenticated to ensure her TEE capability when joining an external network.

Meanwhile, the systems PDOs~\cite{bowman2018private}, Phala~\cite{Phala2019}, Ekiden~\cite{cheng2019ekiden} and COMMITEE~\cite{Erwig2020CommiTEEAE} introduce an expendable and interchangeable solution. TEEs are stateless: any particular TEE can be easily replaced once it has clashed or finished its task. Unfortunately, these solutions are along with new challenges. Firstly, even if TEEs are changeable, detecting a compromised TEE is still difficult. For instance, PDOs can re-execute a method multiple times for the verification. Given the same input parameters to different TEEs, TEEs are believed to work securely only if their outcomes match. Then, the outputs of enclaves are allowed to commit to the blockchain. COMMITEE adopts \textit{master/backup} TEE host mechanism. If the master TEE host is proved to be malicious, a backup TEE host will continue to work without communications to the master TEE host. Nevertheless, this model increases the attack interface and makes the whole system vulnerable. Secondly, TEE hosts are stateless. That means, to ensure an exceptional execution is recoverable, any persistent state must be stored in the blockchain or a trusted third party (TTP). However, for a non-deterministic blockchain system such as Ethereum (PoS version)~\cite{wood2014ethereum}, verifying whether an item has been stored on the blockchain is a non-trivial task. Meanwhile, storing data in TTPs may lead to the single-point failure, which goes against the blockchain's real intention.

\smallskip
\noindent\underline{\textit{TEE security.}} A contract runs inside TEE, and heavily depends on remote attestation service. The SGX-supported blockchain systems including PDOs~\cite{bowman2018private}, Fastkitten~\cite{das2019fastkitten}, ShadowEth~\cite{yuan2018shadoweth}, Phala~\cite{Phala2019} and Ekiden~\cite{cheng2019ekiden} assume that Intel Attestation Service (IAS) is trusted. IAS can correctly and completely report whether a certain output with cryptographic material (\textit{quote}~\cite{costan2016intel}) is produced by SGX-enabled hardware. However, IAS might be compromised, posing a risk to these architectures. A compromised or hijacked remote attestation service may maliciously report an attestation with the wrong cryptographic material that does not belong to its corresponding TEE hardware, breaking the promised security. Meanwhile, a centralized service might be crashed, causing the leakage of private states. Unfortunately, none of layer-two schemes consider these risks in designs or implementations.

As discussed, current TEE implementations have memory limitations for confidential executions. If the memory usage exceeds the threshold, it may confront significant performance and security issues \cite{weichbrodt2018sgx}. Hybridchain~\cite{wang2020hybridchain} optimizes the storage by maintaining transaction records outside Intel SGX. Meanwhile, TZ4Fabric \cite{muller2020tz4fabric} minimizes TCB by avoiding all the executions inside TEEs. However, these approaches increase the implementation complexity. A well-known fact is that a TEE is vulnerable to physical vulnerabilities \cite{hill2019spectre}. Unfortunately, very few layer-two solutions provide remedial measures to reduce the risk of being attacked.

\begin{table*}[h]
\caption{\textbf{Evaluation for current TEE-assisted confidential smart contract systems}}
\label{tab-pitfall}
\begingroup
    \newcommand*{\HeadAux}[1]{%
      \multicolumn{1}{@{}r@{}}{%
        \vcorr
        \sbox0{\headformat{\strut #1}}%
        \sbox2{\headformat{}}%
        \sbox4{\kern\tabcolsep\redtriangle\kern\tabcolsep}%
        \sbox6{\rotatebox{65}{\rule{0pt}{\dimexpr\ht0+\dp0\relax}}}%
        \sbox0{\raisebox{.5\dimexpr\dp0-\ht0\relax}[0pt][0pt]{\unhcopy0}}%
        \kern.75\wd4 %
        \rlap{%
          \raisebox{.25\wd4}{\rotatebox{55}{\unhcopy0}}%
        }%
        \kern.25\wd4 %
        \ifx\HeadLine Y%
          \dimen0=\dimexpr\wd2+.5\wd4\relax
        \fi
      }%
    }%
    \newcommand*{\head}[1]{\HeadAux{\global\let\HeadLine=Y#1}}%
    \newcommand*{\headNoLine}[1]{\HeadAux{\global\let\HeadLine=N#1}}%
    \noindent
\resizebox{0.95\linewidth}{!}{    
    \begin{tabular}{%
      l|>{\quad}c
      *{5}{|c}|c
      *{5}{|c}|c
      *{6}{|c}|c
      *{5}{|c}c
    }%
    \toprule
     \head{\textbf{Selected Examples}} &
      \head{Host incentive mechanism} &
      \head{Host punishment mechanism} &
      \head{Host fault tolerance} &
      \head{Host authentication} &
      \headNoLine{} &
     \head{TEE attestation security} &
      \head{TEE memory limitation} &
      \head{TEE physical attacks} &
      \head{TEE trusted timer} &
      \headNoLine{} &
      \head{Workload Measurement} &
      \head{Flaws detection} &
      \head{User query restriction} &
      \head{Blockchain data confirmation} &
      \head{TEE output conflicts} &
      \headNoLine{} & 
      \head{Distributed key protocol} &
      \head{Key rotation protocol} &
      \head{Independent contract key} &
      \head{Independent TEE key} &
      \headNoLine{} & \headNoLine{}

      \sbox0{S}
    \\ \midrule
     \rowcolor{tabyellow}%
     2017, Enigma~\cite{enigma,TheDevel7enigma} & \redtriangle  & \redtriangle    & \greenbullet &  \greenbullet
     && 
     \redtriangle& \redtriangle& \greenbullet &  \redtriangle 
     && 
     \greenbullet & \redtriangle & \redtriangle & \greenbullet & \greenbullet 
     && 
     \redtriangle & - &  \redtriangle & $\star$ & \multicolumn{2}{c}{}
   
      \\\hline     
      \rowcolor{tabyellow}%
      2018, Fabric \cite{brandenburger2018blockchain}  &  \redtriangle &  \redtriangle & \greenbullet & \greenbullet  
      &&  
      \orangecirc &  \redtriangle  & \greenbullet & \redtriangle 
      && 
      \greenbullet &  \redtriangle &  \redtriangle & \greenbullet & \greenbullet 
      && 
      \redtriangle & \redtriangle & \redtriangle & \redtriangle & \multicolumn{2}{c}{}
      
     \\\hline 
      \rowcolor{tabyellow}%
      2019, CCF \cite{russinovich2019ccf} &  \redtriangle &  \redtriangle & \greenbullet & \greenbullet 
      &&  
      \redtriangle & \redtriangle  & \greenbullet &\redtriangle 
      && 
      \greenbullet & \redtriangle &  \redtriangle &\greenbullet & \greenbullet 
      && 
      \redtriangle  & \greenbullet & \redtriangle & \redtriangle & \multicolumn{2}{c}{}
      
     \\\hline
      \rowcolor{tabyellow}%
      2020, CONFIDE \cite{yan2020confidentiality}  &  \redtriangle &  \redtriangle &  \redtriangle &\greenbullet   
      &&  
      \greenbullet &  \redtriangle  & \redtriangle & \redtriangle 
      &&
      \greenbullet &\redtriangle  &   \redtriangle & \greenbullet  & \greenbullet 
      &&
      \greenbullet & - & \redtriangle & \redtriangle & \multicolumn{2}{c}{}
      
     \\\hline 
     2016, Hawk~\cite{kosba2016hawk}  &  \redtriangle  & \redtriangle & - & -  
     && 
     - & -& \redtriangle & -
     &&
     - & -& - & - & - 
     && 
     - & -  & -  & - & \multicolumn{2}{c}{}
  
     \\\hline 
     2018, PDOs~\cite{bowman2018private}  &  \redtriangle & \redtriangle & \greenbullet & \greenbullet
     &&
     \redtriangle & \redtriangle  & \redtriangle& \redtriangle
     && 
     \redtriangle & \greenbullet & \redtriangle & \redtriangle & \greenbullet 
     && 
     \redtriangle & \redtriangle & \greenbullet  & \greenbullet & \multicolumn{2}{c}{}
     
     \\\hline
     2018, ShadowEth~\cite{yuan2018shadoweth} &  
     \greenbullet  & \redtriangle & \greenbullet & \greenbullet
     &&  
     \redtriangle & \redtriangle  & \redtriangle& \redtriangle 
     &&   
     \greenbullet & \redtriangle &  \greenbullet
     & \greenbullet &  \greenbullet 
     && 
     \redtriangle & \greenbullet & \greenbullet  & \redtriangle  & \multicolumn{2}{c}{}
     
     \\\hline 
     2019, Phala~\cite{Phala2019}  &   \greenbullet  & \redtriangle & \greenbullet & \redtriangle
     && 
     \greenbullet & \redtriangle  & \redtriangle& \redtriangle &  
     &\redtriangle & \redtriangle & \greenbullet & \greenbullet &\greenbullet 
     && 
     \orangecirc & \greenbullet & \greenbullet  & \redtriangle  & \multicolumn{2}{c}{}
     
     \\\hline   
     2019, Ekiden~\cite{cheng2019ekiden} &  \redtriangle & \redtriangle & \greenbullet & \greenbullet
     && 
     \redtriangle & \redtriangle & \orangecirc & \greenbullet 
     && 
     \redtriangle & \redtriangle &   
     \orangecirc &\greenbullet  & \greenbullet 
     && 
     \greenbullet & \greenbullet & \redtriangle  & \redtriangle  & \multicolumn{2}{c}{}
     
     \\\hline
     2019, Fastkitten~\cite{das2019fastkitten} &  \redtriangle & \greenbullet & \redtriangle & \greenbullet
     && 
     \redtriangle  & \redtriangle & \redtriangle & \redtriangle 
     &&  
     \greenbullet & \redtriangle & \greenbullet & \greenbullet & \redtriangle 
     && 
     \orangecirc & \redtriangle & \redtriangle &  \redtriangle & \multicolumn{2}{c}{}
       
    \\\hline
    2019, Avalon~\cite{avalon19} & \redtriangle  & \redtriangle  &  \greenbullet  & \greenbullet 
    && 
     \redtriangle &  \redtriangle & \greenbullet  & \redtriangle
    && 
     \redtriangle & \redtriangle & \redtriangle &  - & - 
    && 
     \redtriangle & \redtriangle &  \redtriangle & \greenbullet &
    \multicolumn{2}{c}{}
    
    \\\hline
    2020, Hybridchain~\cite{wang2020hybridchain} &  \redtriangle & \redtriangle  & \greenbullet & \redtriangle
    && 
    \redtriangle & \greenbullet  & \orangecirc &  \redtriangle
    &&  
    \redtriangle & \redtriangle & \redtriangle  & \redtriangle & \redtriangle
    && 
    \redtriangle & \redtriangle & \redtriangle  & \redtriangle & \multicolumn{2}{c}{}
      
    \\\hline
    2020, COMMITEE~\cite{Erwig2020CommiTEEAE} &  \redtriangle & \redtriangle & \greenbullet  & \greenbullet 
    && 
    \redtriangle & \redtriangle  & \redtriangle & \redtriangle 
    &&   
    \redtriangle & \greenbullet &  \redtriangle & \greenbullet & \greenbullet 
    && 
    \redtriangle & \redtriangle & \redtriangle  & \redtriangle & \multicolumn{2}{c}{}
      
    \\\hline
    2020, PrivacyGuard~\cite{xiao2020privacyguard} &  \greenbullet & \redtriangle & \redtriangle  & \redtriangle 
    && 
    \redtriangle & \redtriangle  & \redtriangle & \redtriangle 
    &&  
    \redtriangle & \redtriangle &  \redtriangle & \greenbullet & \greenbullet 
    && 
    \redtriangle & \redtriangle & \redtriangle  & \redtriangle & \multicolumn{2}{c}{}
      
    \\\hline
     2020, TZ4Fabric~\cite{muller2020tz4fabric} & \redtriangle  & \redtriangle & \redtriangle & \redtriangle
     && 
     -  & \greenbullet & \redtriangle & \redtriangle  
     && 
    \redtriangle & \redtriangle & \greenbullet & \redtriangle & \greenbullet 
     && 
     \redtriangle & \redtriangle & \redtriangle & \greenbullet & \multicolumn{2}{c}{}

    \\\hline
    2020, Taxa~\cite{taxa2020} &  \redtriangle & \greenbullet  & \greenbullet  & \redtriangle 
    && 
    \redtriangle & \redtriangle  & \redtriangle & \redtriangle
    && 
    \redtriangle &\redtriangle & \redtriangle & \redtriangle & \greenbullet 
    && 
    \redtriangle & \redtriangle & \redtriangle  & \redtriangle & \multicolumn{2}{c}{}
    \\\hline
    
    2021, Erdstall~\cite{Introduc42,Technolo53} &  \redtriangle  & \greenbullet  & \greenbullet  & \greenbullet
    && 
    \redtriangle & \redtriangle  & \redtriangle & \redtriangle
    && 
    \redtriangle & \redtriangle & \redtriangle & \redtriangle & \greenbullet 
    && 
    \redtriangle & \redtriangle & \redtriangle  & \greenbullet & 
    \multicolumn{2}{c}{}
      
    \\ \midrule
      \multicolumn{1}{c}{} &
      \multicolumn{5}{c}{\bfseries TEE Host Security} & 
      \multicolumn{5}{c}{\bfseries TEE Security} & 
      \multicolumn{6}{c}{\bfseries TEE Program Security}  &
      \multicolumn{6}{c}{\bfseries TEE Key Security}
      \\
    \bottomrule
    \end{tabular}
    }
    \kern19.5mm 
  \endgroup
  \begin{tablenotes}
       \footnotesize
       \item[1] \greenbullet\, Considered this pitfall with offering remedy; \orangecirc\, Discussed this pitfall without offering remedy; $\star$ Without reference.  \\  \redtriangle\,  Did not consider this pitfall;  $-$\, The system is secure without this pitfall; \, Layer-one solutions are in \colorbox{tabyellow}{\color{black}Yellow} background. 
     \end{tablenotes}
\end{table*}

\smallskip
\noindent\underline{\textit{TEE program security.}} A poorly-written contract might deviate from designated functionalities and further leak the secret information. This part discusses the potential pitfalls and remedies when deploying contracts.

In original smart contract systems, gas mechanism is a powerful tool to prevent \textit{infinite loop} attacks~\cite{wood2014ethereum}. Since the layer-two systems execute smart contract outside the blockchain, a similar mechanism must be considered. Fastkitten~\cite{das2019fastkitten} and Hybridchain~\cite{wang2020hybridchain} protect against such attacks by using the \textit{timeout} mechanism. Limitations are firstly defined on the maximum amount of execution steps that allow to perform inside a TEE per round. Then, TEE monitors smart contract operations. If the number of execution steps exceeds a predefined threshold, the enclave will terminate executions. ShadowEth~\cite{yuan2018shadoweth} combines a timeout mechanism with a \textit{remuneration} mechanism. Similar to the gas mechanism in Ethereum~\cite{wood2014ethereum}, TEE hosts can still gain remuneration even if a contract exits after timeout since they provide sufficient computing power. These mechanisms effectively protect against endless loops and denial-of-service (DoS) launched by external attackers.

The TEE itself lacks self-awareness of input data, since it cannot distinguish which state is fresh. A lack of input data authentication makes the system vulnerable to the rollback attack~\cite{Pries2008ANR,homoliak2020aquareum}. A malicious user may attempt to invoke the confidential contract many times to seek the leaked secret information. Authentication of the user's identity is helpful to prevent this attack. However, none layer-two solution provides these remedies for these potential pitfalls. On the other hand, the TEE input may come from a non-deterministic blockchain system~\cite{brandenburger2019trusted,zhang2019solution}, in which deciding whether an input has been confirmed is tricky. Fastkitten~\cite{das2019fastkitten} and COMMITEE~\cite{Erwig2020CommiTEEAE} mitigate this issue by using a \textit{checkpoint} mechanism. As for TEE output conflicts, Ekiden~\cite{cheng2019ekiden} uses a probabilistic proof-of-publication protocol to avoid the ambiguous input.

After the invocation of a private contract, the outputs returned from TEEs are uploaded on-chain for the final confirmation. But a malicious TEE host may send an exceptional result to the blockchain. Even worse, two hosts may publish different updates towards the same contract simultaneously. To prevent such malicious publications and to evade conflicts, PDOs~\cite{bowman2018private} depends on Coordination and Commit Log (CCL) to manage the synchronization in the execution of interacting contracts and enables a contract owner to decide on selecting the enclave for contract executions, which effectively avoid conflicts. Phala~\cite{Phala2019} adopts an event sourcing command query responsibility segregation architecture to scale up and avoid conflicts, in which the write operations are recorded as events and read operations can be served by the current view of states. Again, these solutions contradict the property of decentralization. Ekiden~\cite{cheng2019ekiden} and ShadowEth~\cite{yuan2018shadoweth} rely on the blockchain to resolve conflicts resulting from concurrency. In particular, ShadowEth~\cite{yuan2018shadoweth} requires a worker to specify the version number with a timestamp when pushing data to the blockchain. Even miners accept different responses at first, they will eventually reach an agreement by comparing version number and the timestamp, with the help of the consensus procedure. Yet, such an approach is inefficient, especially in non-deterministic blockchain systems.

\smallskip
\noindent\underline{\textit{TEEs key management.}} PDOs \cite{bowman2018private} uses a key provisioning service to distribute private keys. The drawback is obvious: A compromised provisioning service could make the entire system fail. To increase the robustness of a private key, Ekiden \cite{cheng2019ekiden} designs a distributed key generation (DKG) \cite{gennaro1999secure} protocol using the secret sharing scheme~\cite{shamir1979share}. Even if one key manager is compromised, an adversary cannot obtain the entire key. However, this solution does not completely solve the key leakage issue. The final keys are assembled and replicated among all end-TEEs. If an adversary compromises an end-TEE, exposing all the contract state becomes a trivial task. The key rotation technology, adopted by Ekiden \cite{cheng2019ekiden}, Fastkitten~\cite{das2019fastkitten}, Phala~\cite{Phala2019} partially solves the above issue by providing a short-term key in every epoch. An adversary cannot corrupt a future or previous committed state, which minimizes the possibility of key exposure to attackers and further helps the layer-two system to achieve forward secrecy. Also, layer-two projects such as COMMITEE~\cite{Erwig2020CommiTEEAE} mitigate these key issues by providing each TEE per secret key. Even if a certain TEE's private key were stolen, this only would affect the smart contract running on that compromised TEE. Furthermore, Phala Network~\cite{Phala2019}, equips each contract with an asymmetric key called the \textit{contract key}, which also enhances the key security to a certain degree.

\subsection{Pros and Cons} 
The layer-two solution decreases computational burden and avoids latency by decoupling the smart contract executions from consensus mechanisms. The solution merely puts the execution results on-chain rather than all processing states. Meanwhile, the layer-two solution does not require a dedicated public ledger, meaning that such a solution can smoothly integrate with existing public blockchain platforms. Unfortunately, this method also brings security and functionality challenges when delegating the task of contract management to an external TEE layer.

\begin{center}
\begin{tcolorbox}[colback=gray!10,
                  colframe=black,
                  width=8cm,
                  arc=1mm, auto outer arc,
                  boxrule=0.5pt,
                 ]

In addition to bringing the privacy properties, the layer-two solution enables complex contract executions without slowing down the consensus process, reducing contract costs and improving performance and scalability. 
\end{tcolorbox}
\end{center}

Firstly, the layer-two solution complexifies contract data management. The contracts that are deployed outside the blockchain require an external execution/storage party. A malicious storage maintainer may reject to provide the service, while a malicious host may abort TEE executions, terminate enclaves or delay/drop messages. Even an honest host might accidentally lose states in a power cycle. To solve the centralization issue and tolerate host failures, many countermeasures such as the TEE network, stateless TEEs and punishment mechanisms, are proposed. However, these solutions are not effortless, inevitably making the system complicated and hard to implement in practice.

Secondly, the layer-two solution increases the attack surface and thus becomes vulnerable to rollback attacks. There is a high probability that an adversary node can revert transactions where temporary forks, representing inconsistent blockchain views, are allowed in blockchain systems with probabilistic consensus (e.g., PoW). Since TEEs provide no guarantee on verification of input data; they cannot distinguish whether an input state is fresh or not. An attacker may offer stale states to resume a TEE’s execution. This enables rollback attacks against randomized TEEs programs. Even worse, plugging up these loopholes needs much effort.

\section{Discussion}
\label{sec-discussion}
This section compares layer-one and layer-two solutions, and discusses the hardware's options and impacts.

\subsection{L1 and L2 Comparison}
\textit{Which solution is more secure?} Even if we have built clear security metrics based on threat models and give concise security analyses in the context of layered architectures, it is still inadequate for answering this question: \textit{Which solution is more secure, layer-one solution or layer-two solution?} This is because system security is a multidimensional topic, and measuring all security aspects is impractical. The security flaws may happen in any phase in a system \cite{pfleeger2010measuring}. Despite some projects performing well in our evaluation, we cannot roughly say that they are more secure. As hybrid technologies, both layer-one and layer-two systems have unsatisfactory security vulnerabilities in existing systems, and they must be carefully treated when applying them to real applications. Frankly speaking, there is a long road to achieving such a practically secure and confidential system. Our aim is not to argue which solution is more secure. Instead, we focus on helping developers and communities to establish a security measurement and avoid potential pitfalls in designing TCSC.

\textit{Which solution is more efficient?} The layer-one solutions require the contract to be confidentially executed in a distributed TEE network, which is time-consuming and hard to scale out. In contrast, layer-two systems only upload final calculated results from offline TEEs to online blockchains. Local TEE hosts can execute complicated computations with high scalability and short execution time. Assuming that the on-chain processing time remains stable, the overall performance gets improved by enabling parallel off-chain executions. Thus, from the view of performance and scalability, the layer-two solution is our recommendation.

\textit{Which solution is more adoptable?} From the aforementioned discussion, we can observe that the layer-one and layer-two solutions fit different scenarios. The layer-one solution is more adoptable in consortium blockchain systems, while the layer-two solution well fits the existing public blockchain systems. Layer-one systems require each blockchain node to equip a TEE, which is difficult to be fulfilled in a public blockchain while already in use. In a consortium blockchain, the nodes are controllable and manageable, and the committee can require each node to equip a TEE when joining the network. On the flip side, the layer-two solution does not change the original blockchain trust assumption. Instead, it creates an independent layer for executing the smart contract, and thus allows developers to seamlessly integrate the TEE into existing public blockchains without significant modifications.

\subsection{Hardware-anchored TEE Options}

Securing smart contracts with TEEs is challenging because we have to assume a strong attacker model, in which the attacker has physical possession of the hardware running the smart contract and can interfere with it in powerful ways. This part discusses the security impact of choosing different TEE architectures. In particular, we select \textit{Intel SGX}~\cite{intel2016}, \textit{Arm TrustZone}~\cite{pinto2019demystifying} and \textit{dedicated chip}~\cite{johnson2018titan} as examples.

Intel SGX is a system allowing one to set up protected enclaves running on an Intel processor. Such enclaves are protected from malware running outside the enclave, including in the operating system. Enclaves can attest their software and computations using a signing key ultimately certified by Intel. Intel SGX has been marketed for desktop machines and servers alike; Microsoft Azure \cite{azure2021} is a commercial cloud offering that allows cloud customers to set up SGX enclaves in the cloud. Many attacks on SGX have been published in the eight years since its release. They may be categorised as side-channel attacks \cite{van2018foreshadow}, fault attacks  \cite{murdock2020plundervolt,chen2021voltpillager} and software attacks \cite{van2019tale}. While some of these attacks can be solved by improvements of SGX, it is unclear that it will ever be possible to have a completely secure version, because the attack surface is large, in the case of smart contracts, one has to assume that attackers have physical possession of the hardware.

ARM TrustZone~\cite{pinto2019demystifying} is a technology widely used in mobile phones to protect secrets, such as secrets used in banking apps. Its ubiquity makes it an attractive option. However, it has been attacked even more than Intel SGX, and doesn’t offer a suitable attestation framework. Future hardware-anchored security products from ARM may address this problem.

Dedicated chips such as the Open Titan~\cite{johnson2018titan} family of chips offer a better solution. Open Titan is an open-source design inspired by Google Titan, a chip used on Google servers and in Google mobile phones. The fact that the smart contract runs on a dedicated chip not shared with attacker code means that the attack surface is much smaller. Attestation frameworks exist for such chips, and the attestation keys can be rooted in a manufacturer’s certificate.  The kind of attacks mentioned for SGX become much harder to mount. Nevertheless, even dedicated chips may succumb to a dedicated and resourceful attacker. Researchers have succeeded in mounting attacks based on power side-channels and electromagnetic (EM) radiation side channels. Defences against such attacks include masking, which consists of randomly splitting every sensitive intermediate variable into multiple shares. Even if the adversary is able to learn a share of the secret via side-channel, it would need all of them in order to recover the secret. Fault attacks such as EM and voltage glitching are also possible, but again, there are known defences \cite{biham1997differential} at both a software and hardware level. Software defences include making secret-dependent computations twice (in general $n$ times) and then comparing results before producing any output. Countermeasures in hardware involve having internal voltage monitoring circuitry, which makes sure that the input voltage remains within a safe operation range and resets the device otherwise. 

\section{Research Challenges} 
\label{sec-chall}

\noindent\textbf{Side channel attack.} Inevitably, all types of TEEs suffer from side-channel attacks. An attacker may observe untrusted resources to obtain the control flow and data access mode from the running hardware to infer sensitive information. Beyond the normal side-channel attack, an attacker can keep track of the changes in encrypted states recorded on the blockchain to extract secrets. The attacker carefully compares encrypted states before and after running a particular confidential transaction. Even if the attacker cannot directly learn about the plaintext, the changes of the encrypted state may lead to a valuable side-channel attack. Come back to the e-voting example, the changes of state $c_{b}'$ indicates a specific sender or receiver's invocation, and ciphertext length reveals which method is being invoked given different arguments size. Meanwhile, the application binary interface (ABI)~\cite{wood2014ethereum}, and the contract path will be spied by an attacker, causing data leakage.

\smallskip
\noindent\textbf{Key management dilemma.} The private keys in TEE-assisted systems are extremely crucial but hard to manage. On the one hand, putting the application keys in a single TEE contributes to the key security. However, it also makes the system raise the risk of a single point of failure. On the other hand, sharing the private key among multiple TEEs offers practical availability but (as a sacrifice) increases key exfiltration risk. Meanwhile, key sharing technologies are too complicated to adopt and cannot completely solve the key issues. Suppose that an attacker steals the attestation key somehow. She might consequently generate the attestation materials to deceive the user with a fake fact: The contract has been executed. Even worse, if a root key stored in the tamper-resistant hardware (e.g., Memory Encryption Engine Key in SGX) is compromised, all key technologies for protecting application keys become useless.

\medskip
\noindent\textbf{Transparency issues.} Compared with cryptographic approaches backed by mathematics~\cite{kalodner2018arbitrum,bunz2018bulletproofs,zyskind2015enigma}, the confidential smart contracts relied on TEEs are lack of transparency. On the one hand, contracts are executed inside TEEs, and the outputs are usually encrypted, which lacks public verifiability inherited from traditional blockchain systems. The attestation service can only guarantee that the encrypted outputs indeed come from a TEE. However, neither users nor the blockchain nodes can learn whether a TEE is compromised or executes contracts following the predefined specifications. Even if many TEEs can re-execute the same contract with the same setup (e.g., the same private key) to check outputs, this inevitably increases the key exfiltration risk in the face of a confidentiality breach. On the other hand, the precise architectures of chips are still unclear for some TEE products, such as Intel SGX~\cite{costan2016intel}. TEE-assisted solutions force the user to put too much trust in the manufacturers of this hardware. Users even argue that Intel may have reduced the security of SGX to improve performance to cater for market demand~\cite{dinh2019everything}. Additionally, the attestation service used to prove that a program runs inside TEEs is \textit{centralized} and \textit{non-transparent}. A compromised provider has the ability to insert fake IDs, and further, steal the confidential state in smart contracts.

\section{Concluding Remarks}
\label{sec-conclu}
The technologies on how to combine smart-contract execution with TEEs are mushrooming nowadays. The absence of systematic work confuses newcomers. In this paper, we provide the first SoK on TEE-assisted confidential smart contract systems. TEE technologies empower transparent smart contracts with confidentiality, greatly extending the scope of upper-layer applications. We summarize state-of-the-art solutions by proposing a unified framework covering aspects of design models, desired properties, and security considerations. Our analysis clarifies existing challenges and future directions for two mainstream architectures (layer-one and layer-two solutions). We believe that this work represents a snapshot of the technologies that have been open-sourced and made public in time. Our evaluation and analysis within this SoK will offer a good guide for communities, and greatly promote the prosperity of development for TCSC applications.

\smallskip
\noindent\textbf{Acknowledgement.} Rujia Li and Qi Wang are partially supported by the Shenzhen Fundamental Research Programs under Grant No.20200925154814002. We thank Xinrui Zhang (SUSTech) for her help. Also, we express our appreciation to anonymous reviewers for their valuable comments.

\normalem
\bibliographystyle{unsrt}
\bibliography{bib}

\begin{thebibliography}{100}

\bibitem{szabo1996smart}
Nick Szabo.
\newblock Formalizing and securing relationships on public networks.
\newblock {\em First monday}, 1997.

\bibitem{wood2014ethereum}
Gavin Wood et~al.
\newblock Ethereum: A secure decentralised generalised transaction ledger.
\newblock {\em \url{https://ethereum.github.io/yellowpaper/paper.pdf}}, 2022.

\bibitem{delmolino2016step}
Kevin Delmolino et~al.
\newblock Step by step towards creating a safe smart contract: Lessons and
  insights from a cryptocurrency lab.
\newblock In {\em FC}, pages 79--94. Springer, 2016.

\bibitem{hewa2021survey}
Hewa et~al.
\newblock Survey on blockchain based smart contracts: Technical aspects and
  future research.
\newblock {\em IEEE Access}, 2021.

\bibitem{alharby2017blockchain}
Maher Alharby and Aad Van~Moorsel.
\newblock Blockchain-based smart contracts: A systematic mapping study.
\newblock {\em arXiv preprint arXiv:1710.06372}, 2017.

\bibitem{jansen2019smart}
Marc Jansen et~al.
\newblock Do smart contract languages need to be turing complete?
\newblock In {\em CBA}, pages 19--26. Springer, 2019.

\bibitem{raval2016decentralized}
Siraj Raval.
\newblock {\em Decentralized applications: harnessing Bitcoin's blockchain
  technology}.
\newblock " O'Reilly Media, Inc.", 2016.

\bibitem{zou2019smart}
Weiqin Zou et~al.
\newblock Smart contract development: Challenges and opportunities.
\newblock {\em TSE}, 2019.

\bibitem{zhang2019security}
Rui Zhang, Rui Xue, and Ling Liu.
\newblock Security and privacy on blockchain.
\newblock {\em CSUR}, 52(3):1--34, 2019.

\bibitem{Goldfeder2018PrivateSC}
Steven Goldfeder.
\newblock Private smart contracts.
\newblock 2018.

\bibitem{steffen2019zkay}
Samuel S., Benjamin Bichsel, Mario Gersbach, Noa Melchior, Petar Tsankov, and
  Martin Vechev.
\newblock zkay: Specifying and enforcing data privacy in smart contracts.
\newblock In {\em CCS}, pages 1759--1776, 2019.

\bibitem{baghery2019efficiency}
Karim Baghery.
\newblock On the efficiency of privacy-preserving smart contract systems.
\newblock In {\em AFRICACRYPT}, pages 118--136. Springer, 2019.

\bibitem{Unterweger2018LessonsLF}
A.~Unterweger, F.~Knirsch, et~al.
\newblock Lessons learned from implementing a privacy-preserving smart contract
  in ethereum.
\newblock {\em NTMS}, pages 1--5, 2018.

\bibitem{zhang2016town}
Fan Zhang, Ethan Cecchetti, Kyle Croman, Ari Juels, and Elaine Shi.
\newblock Town crier: An authenticated data feed for smart contracts.
\newblock In {\em CCS}, pages 270--282, 2016.

\bibitem{blass2019borealis}
Erik-Oliver Blass and Florian Kerschbaum.
\newblock Borealis: Building block for sealed bid auctions on blockchains.
\newblock In {\em AsiaCCS}, pages 558--571, 2020.

\bibitem{galal2019trustee}
Hisham~S Galal and Amr~M Youssef.
\newblock Trustee: full privacy preserving vickrey auction on top of ethereum.
\newblock In {\em FC}, pages 190--207. Springer, 2019.

\bibitem{cortier2016sok}
V{\'e}ronique Cortier, David Galindo, Ralf K{\"u}sters, Johannes Mueller, and
  Tomasz Truderung.
\newblock Sok: Verifiability notions for e-voting protocols.
\newblock In {\em SP}, pages 779--798. IEEE, 2016.

\bibitem{Rathee2021OnTD}
Geetanjali Rathee et~al.
\newblock On the design and implementation of a blockchain enabled e-voting
  application within iot-oriented smart cities.
\newblock {\em IEEE Access}, 9:34165--34176, 2021.

\bibitem{gdpr}
General data protection regulation. \url{https://gdpr-info.eu/}.
\newblock 2020.

\bibitem{voigt2017eu}
Paul Voigt et~al.
\newblock The eu general data protection regulation (gdpr).
\newblock {\em A Practical Guide, 1st Ed., Cham: Springer International
  Publishing}, 10:3152676, 2017.

\bibitem{kosba2016hawk}
Ahmed Kosba, Andrew Miller, Elaine Shi, Zikai Wen, and Charalampos Papamanthou.
\newblock Hawk: The blockchain model of cryptography and privacy-preserving
  smart contracts.
\newblock In {\em SP}, pages 839--858. IEEE, 2016.

\bibitem{kalodner2018arbitrum}
Harry Kalodner et~al.
\newblock Arbitrum: Scalable, private smart contracts.
\newblock In {\em USENIX Security}, pages 1353--1370, 2018.

\bibitem{bunz2018bulletproofs}
B.~B{\"u}nz, Jonathan Bootle, Dan Boneh, Andrew Poelstra, Pieter Wuille, and
  Greg Maxwell.
\newblock Bulletproofs: Short proofs for confidential transactions and more.
\newblock In {\em SP}, pages 315--334. IEEE, 2018.

\bibitem{bunz2020zether}
Benedikt B{\"u}nz et~al.
\newblock Zether: Towards privacy in a smart contract world.
\newblock In {\em FC}, pages 423--443. Springer, 2020.

\bibitem{chen2020pgc}
Yu~Chen, Xuecheng Ma, Cong Tang, and Man~Ho Au.
\newblock Pgc: Decentralized confidential payment system with auditability.
\newblock In {\em ESORICS}, pages 591--610. Springer, 2020.

\bibitem{solomon2021smartfhe}
Ravital Solomon et~al.
\newblock smartfhe: Privacy-preserving smart contracts from fully homomorphic
  encryption.
\newblock {\em IACR Cryptol. ePrint Arch.}, 2021:133, 2021.

\bibitem{zyskind2015enigma}
Guy Zyskind et~al.
\newblock Enigma: Decentralized computation platform with guaranteed privacy.
\newblock {\em arXiv:1506.03471}, 2015.

\bibitem{lee2020keystone}
Dayeol Lee, David Kohlbrenner, et~al.
\newblock Keystone: An open framework for architecting trusted execution
  environments.
\newblock In {\em EuroSys}, pages 1--16, 2020.

\bibitem{ekberg2013trusted}
Jan-Erik Ekberg et~al.
\newblock Trusted execution environments on mobile devices.
\newblock In {\em CCS}, pages 1497--1498, 2013.

\bibitem{kim2017enhancing}
Seongmin Kim et~al.
\newblock Enhancing security and privacy of tor's ecosystem by using trusted
  execution environments.
\newblock In {\em NSDI}, pages 145--161, 2017.

\bibitem{kaplan2016amd}
David Kaplan, Jeremy Powell, and Tom Woller.
\newblock Amd memory encryption.
\newblock {\em White paper}, 2016.

\bibitem{brasser2019sanctuary}
Ferdinand Brasser, David Gens, Patrick Jauernig, Ahmad-Reza Sadeghi, and
  Emmanuel Stapf.
\newblock Sanctuary: Arming trustzone with user-space enclaves.
\newblock In {\em NDSS}, 2019.

\bibitem{mckeen2013innovative}
Frank McKeen, Ilya Alexandrovich, Alex Berenzon, Carlos~V Rozas, Hisham Shafi,
  Vedvyas Shanbhogue, and Uday~R Savagaonkar.
\newblock Innovative instructions and software model for isolated execution.
\newblock {\em Hasp@ isca}, 10(1), 2013.

\bibitem{zhao2016performance}
ChongChong Zhao et~al.
\newblock On the performance of intel sgx.
\newblock In {\em WISA}, pages 184--187. IEEE, 2016.

\bibitem{cui2021dynamic}
Jinhua Cui et~al.
\newblock Dynamic binary translation for sgx enclaves.
\newblock {\em arXiv preprint arXiv:2103.15289}, 2021.

\bibitem{li2021offline}
Rujia Li, Qin Wang, et~al.
\newblock An offline delegatable cryptocurrency system.
\newblock {\em arXiv preprint arXiv:2103.12905}, 2021.

\bibitem{yan2020confidentiality}
Ying Yan, Changzheng Wei, et~al.
\newblock Confidentiality support over financial grade consortium blockchain.
\newblock In {\em SIGMOD}, pages 2227--2240, 2020.

\bibitem{sinhaluciditee}
Rohit Sinha et~al.
\newblock Luciditee: A tee-blockchain system for policy-compliant multiparty
  computation with fairness.

\bibitem{unlockblockchain_2021}
Chinese chang'an chain enterprise blockchain joins digital yuan project, Mar
  2021.

\bibitem{financials21}
Financials.
\newblock Changan chain, the first independent and controllable blockchain
  technology system in china, was released today.

\bibitem{wang2020hybridchain}
Yong Wang et~al.
\newblock Hybridchain: A novel architecture for confidentiality-preserving and
  performant permissioned blockchain using trusted execution environment.
\newblock {\em IEEE Access}, 8:190652--190662, 2020.

\bibitem{cheng2019ekiden}
Raymond Cheng, Fan Zhang, Jernej Kos, Warren He, Nicholas Hynes, Noah Johnson,
  Ari Juels, Andrew Miller, and Dawn Song.
\newblock Ekiden: A platform for confidentiality-preserving, trustworthy, and
  performant smart contracts.
\newblock In {\em EuroSP}, pages 185--200. IEEE, 2019.

\bibitem{das2019fastkitten}
Poulami Das et~al.
\newblock Fastkitten: Practical smart contracts on bitcoin.
\newblock In {\em USENIX Security}, pages 801--818, 2019.

\bibitem{muller2020tz4fabric}
Christina M{\"u}ller, Marcus Brandenburger, et~al.
\newblock Tz4fabric: Executing smart contracts with arm trustzone.
\newblock {\em arXiv preprint arXiv:2008.11601}, 2020.

\bibitem{russinovich2019ccf}
Mark Russinovich et~al.
\newblock Ccf: A framework for building confidential verifiable replicated
  services.
\newblock Technical Report MSR-TR-2019-16, Microsoft, April 2019.

\bibitem{bowman2018private}
Mic Bowman et~al.
\newblock Private data objects: an overview.
\newblock {\em arXiv preprint arXiv:1807.05686}, 2018.

\bibitem{young1996dark}
Adam Young and Moti Yung.
\newblock The dark side of “black-box” cryptography or: Should we trust
  capstone?
\newblock In {\em CRYPTO}, pages 89--103. Springer, 1996.

\bibitem{li2019auditable}
Rujia Li, David Galindo, and Qi~Wang.
\newblock Auditable credential anonymity revocation based on privacy-preserving
  smart contracts.
\newblock In {\em CBT}, pages 355--371. Springer, 2019.

\bibitem{li2020accountable}
Rujia Li, Qin Wang, et~al.
\newblock An accountable decryption system based on privacy-preserving smart
  contracts.
\newblock In {\em ISC}, pages 372--390. Springer, 2020.

\bibitem{oasislab}
Oasis lab.
\newblock {\em
  \url{https://github.com/oasislabs/secret-ballot/blob/master/contracts/SecretBallot.sol}}.

\bibitem{cortier2014election}
V{\'e}ronique Cortier et~al.
\newblock Election verifiability for helios under weaker trust assumptions.
\newblock In {\em ESORICS}, pages 327--344. Springer, 2014.

\bibitem{Unger2015SoKSM}
Nik Unger, Sergej Dechand, Joseph Bonneau, Sascha Fahl, H.~Perl, I.~Goldberg,
  and M.~Smith.
\newblock Sok: Secure messaging.
\newblock {\em SP}, pages 232--249, 2015.

\bibitem{androulaki2013evaluating}
Elli Androulaki, Ghassan~O Karame, Marc Roeschlin, Tobias Scherer, and Srdjan
  Capkun.
\newblock Evaluating user privacy in bitcoin.
\newblock In {\em FC}, pages 34--51. Springer, 2013.

\bibitem{meiklejohn2013fistful}
Sarah Meiklejohn, Marjori Pomarole, Grant Jordan, et~al.
\newblock A fistful of bitcoins: characterizing payments among men with no
  names.
\newblock In {\em IMC}, pages 127--140, 2013.

\bibitem{brasser2017software}
Ferdinand Brasser et~al.
\newblock Software grand exposure:$\{$SGX$\}$ cache attacks are practical.
\newblock In {\em WOOT}, 2017.

\bibitem{xu2015controlled}
Yuanzhong Xu et~al.
\newblock Controlled-channel attacks: Deterministic side channels for untrusted
  operating systems.
\newblock In {\em SP}, pages 640--656. IEEE, 2015.

\bibitem{hill2019spectre}
Mark~D Hill et~al.
\newblock On the spectre and meltdown processor security vulnerabilities.
\newblock {\em IEEE Micro}, 39(2):9--19, 2019.

\bibitem{dwork2008differential}
Cynthia Dwork.
\newblock Differential privacy: A survey of results.
\newblock In {\em International conference on theory and applications of models
  of computation}, pages 1--19. Springer, 2008.

\bibitem{homoliak2020aquareum}
Ivan Homoliak and Pawel Szalachowski.
\newblock Aquareum: A centralized ledger enhanced with blockchain and trusted
  computing.
\newblock {\em arXiv preprint arXiv:2005.13339}, 2020.

\bibitem{brandenburger2018blockchain}
Marcus Brandenburger et~al.
\newblock Blockchain and trusted computing: Problems, pitfalls, and a solution
  for hyperledger fabric.
\newblock {\em arXiv preprint arXiv:1805.08541}, 2018.

\bibitem{enigma}
Enigma – securing the decentralized web.
\newblock {\em \url{https://www.enigma.co/}}.

\bibitem{garay2015bitcoin}
Juan Garay et~al.
\newblock The bitcoin backbone protocol: Analysis and applications.
\newblock In {\em EUROCRYPT}, pages 281--310. Springer, 2015.

\bibitem{garay2017bitcoin}
Juan Garay et~al.
\newblock The bitcoin backbone protocol with chains of variable difficulty.
\newblock In {\em CRYPTO}, pages 291--323. Springer, 2017.

\bibitem{pass2017analysis}
Rafael Pass, Lior Seeman, and Abhi Shelat.
\newblock Analysis of the blockchain protocol in asynchronous networks.
\newblock In {\em EUROCRYPT}, pages 643--673. Springer, 2017.

\bibitem{garay2020sok}
Juan Garay and Aggelos Kiayias.
\newblock Sok: A consensus taxonomy in the blockchain era.
\newblock In {\em RSA}, pages 284--318. Springer, 2020.

\bibitem{SGX2020}
Intel.
\newblock Intel software guard extensions (intel sgx).
\newblock {\em Accessible on
  \url{https://software.intel.com/content/www/us/en/develop/topics/software-guard-extensions.html}},
  2020.

\bibitem{krahn2020teemon}
Robert Krahn, Donald Dragoti, Franz Gregor, et~al.
\newblock Teemon: A continuous performance monitoring framework for tees.
\newblock In {\em Middleware}, pages 178--192, 2020.

\bibitem{yuan2018shadoweth}
Rui Yuan et~al.
\newblock Shadoweth: Private smart contract on public blockchain.
\newblock {\em JCST}, 33(3):542--556, 2018.

\bibitem{Phala2019}
Yin Hang, Zhou Shunfan, and Jiang Jun.
\newblock Phala network: A confidential smart contract network based on
  polkadot.
\newblock {\em https://files.phala.network/phala-paper.pdf}, 2019.

\bibitem{taxa2020}
Taxa.
\newblock Taxa network: a universal logic layer for blockchain.
\newblock Website, 2021.
\newblock \url{https://taxa.network/}.

\bibitem{TheDevel7enigma}
Enigma.
\newblock The developer quickstart guide to enigma | by enigma project |
  enigma.
\newblock {\em
  \url{https://blog.enigma.co/the-developer-quickstart-guide-to-enigma-880c3fc4308}}.

\bibitem{avalon19}
Hyperledger.
\newblock Introducing hyperledger avalon.
\newblock
  \url{www.hyperledger.org/blog/2019/10/03/introducing-hyperledger-avalon},
  2019.
\newblock (Accessed on 04/19/2021).

\bibitem{Erwig2020CommiTEEAE}
Andreas Erwig, S.~Faust, et~al.
\newblock Commitee: An efficient and secure commit-chain protocol using tees.
\newblock {\em IACR Cryptol. ePrint Arch.}, 2020:1486, 2020.

\bibitem{xiao2020privacyguard}
Yang Xiao et~al.
\newblock Privacyguard: Enforcing private data usage control with blockchain
  and attested off-chain contract execution.
\newblock In {\em ESORICS}, pages 610--629. Springer, 2020.

\bibitem{Introduc42}
Perun Network.
\newblock Introducing erdstall: Scaling ethereum using trusted execution
  environments | by perun network | perunnetwork | medium.

\bibitem{Technolo53}
Erdstall.
\newblock Technology – erdstall.
\newblock \url{https://erdstall.dev/technology/}.
\newblock (Accessed on 04/17/2021).

\bibitem{liu2009research}
Wentao Liu.
\newblock Research on dos attack and detection programming.
\newblock In {\em Third International Symposium on Intelligent Information
  Technology Application}, volume~1, pages 207--210. IEEE, 2009.

\bibitem{de2000revisiting}
Roberto De~Prisco et~al.
\newblock Revisiting the paxos algorithm.
\newblock {\em Theoretical Computer Science}, 243(1-2):35--91, 2000.

\bibitem{gavzi2019proof}
Peter Ga{\v{z}}i, Aggelos Kiayias, and Dionysis Zindros.
\newblock Proof-of-stake sidechains.
\newblock In {\em SP}, pages 139--156. IEEE, 2019.

\bibitem{costan2016intel}
Victor Costan and Srinivas Devadas.
\newblock Intel sgx explained.
\newblock {\em IACR Cryptol. ePrint Arch.}, 2016(86):1--118, 2016.

\bibitem{weichbrodt2018sgx}
Nico W., Pierre-Louis Aublin, and R{\"u}diger Kapitza.
\newblock sgx-perf: A performance analysis tool for intel sgx enclaves.
\newblock In {\em Middleware}, pages 201--213, 2018.

\bibitem{Pries2008ANR}
R.~Pries et~al.
\newblock A new replay attack against anonymous communication networks.
\newblock {\em ICC}, pages 1578--1582, 2008.

\bibitem{brandenburger2019trusted}
Marcus Brandenburger, Christian Cachin, R{\"u}diger Kapitza, and Alessandro
  Sorniotti.
\newblock Trusted computing meets blockchain: Rollback attacks and a solution
  for hyperledger fabric.
\newblock In {\em SRDS}, pages 324--32409. IEEE, 2019.

\bibitem{zhang2019solution}
Shenbin Zhang et~al.
\newblock A solution for the risk of non-deterministic transactions in
  hyperledger fabric.
\newblock In {\em ICBC}, pages 253--261. IEEE, 2019.

\bibitem{gennaro1999secure}
Rosario Gennaro, Stanis{\l}aw Jarecki, Hugo Krawczyk, and Tal Rabin.
\newblock Secure distributed key generation for discrete-log based
  cryptosystems.
\newblock In {\em EUROCRYPT}, pages 295--310. Springer, 1999.

\bibitem{shamir1979share}
Adi Shamir.
\newblock How to share a secret.
\newblock {\em Communications of the ACM}, 22(11):612--613, 1979.

\bibitem{pfleeger2010measuring}
Shari Pfleeger and Robert Cunningham.
\newblock Why measuring security is hard.
\newblock {\em IEEE SP}, 8(4):46--54, 2010.

\bibitem{intel2016}
Intel.
\newblock Introduction to intel® sgx sealing.
\newblock Website, 2016.
\newblock
  \url{https://software.intel.com/content/www/us/en/develop/blogs/introduction-to-intel-sgx-sealing.html}.

\bibitem{pinto2019demystifying}
Sandro Pinto and Nuno Santos.
\newblock Demystifying arm trustzone: A comprehensive survey.
\newblock {\em CSUR}, 51(6):1--36, 2019.

\bibitem{johnson2018titan}
Scott Johnson et~al.
\newblock Titan: enabling a transparent silicon root of trust for cloud.
\newblock In {\em Hot Chips: A Symposium on High Performance Chips}, volume
  194, 2018.

\bibitem{azure2021}
Cynthia Dwork.
\newblock Microsoft azure.
\newblock 2021.

\bibitem{van2018foreshadow}
Jo~Van~Bulck et~al.
\newblock Foreshadow: Extracting the keys to the intel sgx kingdom with
  transient out-of-order execution.
\newblock In {\em USENIX Security}, pages 991--1008, 2018.

\bibitem{murdock2020plundervolt}
Kit Murdock, David Oswald, Flavio~D Garcia, et~al.
\newblock Plundervolt: Software-based fault injection attacks against intel
  sgx.
\newblock In {\em SP}, pages 1466--1482. IEEE, 2020.

\bibitem{chen2021voltpillager}
Zitai Chen et~al.
\newblock Voltpillager: Hardware-based fault injection attacks against intel
  sgx enclaves using the svid voltage scaling interface.
\newblock In {\em USENIX Security}, 2021.

\bibitem{van2019tale}
Jo~Van~Bulck, David Oswald, et~al.
\newblock A tale of two worlds: Assessing the vulnerability of enclave
  shielding runtimes.
\newblock In {\em CCS}, pages 1741--1758, 2019.

\bibitem{biham1997differential}
Eli Biham and Adi Shamir.
\newblock Differential fault analysis of secret key cryptosystems.
\newblock In {\em CRYPOTO}, pages 513--525. Springer, 1997.

\bibitem{dinh2019everything}
Tu~Dinh~Ngoc, Bao Bui, et~al.
\newblock Everything you should know about intel sgx performance on virtualized
  systems.
\newblock {\em POMACS}, 3(1):1--21, 2019.

\bibitem{narayanan2016bitcoin}
Arvind Narayanan, Joseph Bonneau, Edward Felten, Andrew Miller, and Steven
  Goldfeder.
\newblock {\em Bitcoin and cryptocurrency technologies: a comprehensive
  introduction}.
\newblock Princeton University Press, 2016.

\bibitem{yuen2020ringct}
Tsz~Hon Yuen, Shi-feng Sun, et~al.
\newblock Ringct 3.0 for blockchain confidential transaction: Shorter size and
  stronger security.
\newblock In {\em FC}, pages 464--483. Springer, 2020.

\bibitem{nakamoto2008bitcoin}
Satoshi Nakamoto.
\newblock Bitcoin: A peer-to-peer electronic cash system.
\newblock Technical report, Manubot, 2008.

\bibitem{lan2021trustcross}
Ying Lan et~al.
\newblock Trustcross: Enabling confidential interoperability across blockchains
  using trusted hardware.
\newblock {\em arXiv preprint arXiv:2103.13809}, 2021.

\end{thebibliography}

\section*{Appendix A. Key Management}
\label{app-a}

A variety of different keys are used in the life cycle of TCSC. For simplicity, we use Intel SGX as the instance. We classify these keys into two types, namely, \textit{service keys} (top half) and \textit{SGX internal keys} (bottom half). 

\smallskip
\noindent\textbf{Service keys.} The keys $sk_{tx}$ and $pk_{m}$ are used to sign a transaction and encrypt a message resulting from a TEE. Correspondingly, the keys $pk_{tx}$ and $sk_{m}$ are used to verify a signature and decrypt a ciphertext, respectively. Meanwhile, the TEE service key $key_{tee}$ is used to encrypt the contract state, and the asymmetric TEE service key $sk_{tee}$ is used to encrypt a voter's input. Since the key management technologies have significant impacts on these service keys, we emphasize them with the yellow background.

\smallskip
\noindent\textbf{SGX internal keys.} The MEE key is generated at boot, and is placed in special registers, and destroyed at system reset. The MEE key is used for memory encryption and decryption, which plays a crucial role in protecting the confidentiality and integrity of enclaves. At the same time, different enclaves in the same TEE platform share one function key, such as the report key and the attestation key~\cite{intel2016}.

\section*{Appendix B. Anonymity and Confidentiality}
\label{app-c}

Anonymity refers to the privacy that relates to real entities, especially for users' identities. In a blockchain system, anonymity indicates that users' transaction activities will not expose any personal information about them. Alternatively, an attack cannot obtain the correct links between real users and their corresponding account/address that sends the transaction~\cite{narayanan2016bitcoin}. Bitcoin and Ethereum only provide a very early version of anonymity, using the pseudonym-based address mechanism to protect identities. However, this cannot guarantee anonymity because attackers can effortlessly map virtual addresses to physical entities through the relationship analysis.

Confidentiality in a blockchain system mainly refers to the privacy of data and contents recorded on-chain~\cite{yuen2020ringct,zhang2019security}. Classic blockchain systems expose all transactions (includes amount information, addresses, amount, etc.) plainly where anyone can read and access. Sensitive information might unconsciously be leaked to malicious analyzers. For instance, ERC20 tokens in the Ethereum system do not provide confidentiality, since anyone can observe every amount's balance. Adversaries can keep tracing the accounts that have a huge amount of tokens and launch attacks such as using the phishing website or cheating through offline activities.

\begin{table*}[!hpt]
\centering
\caption{Key Types in Confidential Smart Contracts: The table shows a voting example achieved by Intel SGX and Ethereum.}\label{tab-keys}
\resizebox{\linewidth}{!}{ 
\begin{tabular}{cp{5.8cm}p{8.5cm}}
\toprule 
\textbf{Keys} & \multicolumn{1}{c}{\textbf{Purpose}} & \multicolumn{1}{c}{\textbf{Remarks}} \\
\midrule
\textit{Transaction signing key $(sk_{tx},vk_{tx})$ } & signing a transaction  & generated by a voter \\
\textit{A vote's message key $(sk_{m},pk_{m})$} & encrypt a message that comes from a voter  & generated by a voter \\
\rowcolor{tabyellow} \textit{TEE service key $key_{tee}$} & encrypt and decrypt a blockchain state  & generated inside an enclave \\
\rowcolor{tabyellow} \textit{TEE service key $(sk_{tee},pk_{tee})$} &  decrypt a user's input/encrypt a TEE's output  & generated inside an enclave \\
\midrule 
\rowcolor{maroon!10} \textit{Memory Encryption Engine Key}  &  encrypt (decrypt) the data before writing (reading) it to (from) RAM. & stored inside a CPU; different enclaves in the same TEE platform share the same MEE key. \\
\rowcolor{maroon!10} \textit{Report key}  & generate a MAC tag for the measurement.
& generated by $\mathsf{EGETKEY}$ instruction; different enclaves in the same TEE platform share one report key;\\  
\rowcolor{maroon!10} \textit{Attestation key}  & produce attestation signatures. & stored in tamper-resistant hardware; different enclaves in the same TEE platform share one attestation key;\\ 
\rowcolor{maroon!10} \textit{Sealing key}  &  migrate secrets between enclaves. & stored in tamper-resistant hardware; different enclaves in the same TEE platform may share the sealing key depending on key policies. \\ 
\bottomrule
\end{tabular}
}
\end{table*}

\section*{Appendix C. Background}
\label{app-c}
\noindent\textbf{Blockchain Technology.} Blockchain, conceptualized by Nakamoto \cite{nakamoto2008bitcoin}, was proposed as a distributed and append-only ledger, in which all committed transactions are stored in a chain of data records (named as blocks). According to the initial idea of Bitcoin~\cite{szabo1996smart}, when the blockchain maintainers reach an agreement on the newest block, related transactions appearing in that time will be packaged in this block and further stored in a distributed network to maintain a continuously growing list. By providing a secure solution to distribute the information and allowing all participants to audit the shared records, blockchain obtains many key characteristics such as decentralization, auditability and non-repudiation, transparency, and non-equivocation.

\smallskip
\noindent\textbf{Smart Contract} Proposed by Szabo~\cite{szabo1996smart}, the smart contract are widely applied in blockchain systems by Ethereum~\cite{wood2014ethereum}. Blockchain-based smart contracts adopt Turing-complete scripting languages to achieve complicated functionalities~\cite{jansen2019smart} and execute thorough state transition/replication over consensus algorithms to realize final consistency. By the design, a blockchain-based smart contract includes multiple functions, methods, and a few parameters that can run on the blockchain when specific conditions or events are met and encompass business logic and transactions between two or more parties. To be specific, the source code of a contract forming as part of a transaction is first sent to the blockchain. Once the transaction is included in a new block and confirmed by the majority of the participants, the contract code becomes immutable and executable. When an external user invokes the contract, the state will be updated under the instruction of the preloaded source code. The neutrality of the execution environment among all blockchain nodes facilitates the same execution result of the program code. Smart contracts thus enable unfamiliar and distributed participants to fairly exchange without trusted third parties and present a uniform approach to improve applications across a wide range of industries.

\smallskip
\noindent\textbf{Trusted Execution Environments.} Trusted Execution Environment (TEE)~\cite{ekberg2013trusted} provides a protected processing area in the main processor that runs on a separation kernel to ensure confidentiality and integrity of inside data and computations. State-of-the-art implementations include Intel Software Guard Extensions (SGX)~\cite{costan2016intel}, ARM TrustZone~\cite{pinto2019demystifying}, Keystone~\cite{lee2020keystone}, \textit{etc}. For a TEE, three main TEE features are highlighted, including \textit{runtime isolation}, \textit{sealing technologies} and \textit{attestation technologies}. For simplicity, we use Intel SGX as an example to explain these features in the following paragraphs. It has to be mentioned that the Intel SGX design used in our paper can also be implemented on other trusted hardware platforms such as Keystone~\cite{lee2020keystone}.

\smallskip
\noindent\hangindent 1em\textit{Runtime Isolation.} The secure and isolated regions of memory are called \textit{enclaves}. Sensitive data and intermediate computations run inside enclaves to provide protection against outside programs. Besides, all the runtime enclave memories are stored in Enclave Page Cache (EPC)~\cite{lan2021trustcross} and encrypted by Memory Encryption Engine (MEE). These protective mechanisms enforced in SGX protect memories against the access of any process outside the enclave itself, including the operating system, hypervisors, etc.

\smallskip
\noindent\hangindent 1em\textit{Sealing.} Sealing~\cite{intel2016} is a process of loading enclave internal secret state to persistent storage. Roughly speaking, using the Sealing, the secrets can be encrypted and stored in the untrusted memory or disk. Further, it allows such encrypted secrets to be retrieved once the enclave is torn down (either due to the host's power or the application itself). Sealing is achieved by using a private seal key~\cite{costan2016intel}, which covers two types of identities: Enclave Identity and Signing Identity. Enclave Identity is represented by the value of \textit{MRENCLAVE}, which is a cryptographic hash of the enclave measurement. Any operation inside an enclave that changes measurement will yield a different key. Thus, it restricts the permission to sealed data; only the corresponding enclave can access the sealed data. In contrast, Signing Identity is provided by an authority and represented by \textit{MRSIGNER}. It provides the same sealing key for different enclaves, or even different versions of the same enclave. Therefore, Signing Identity can be used to share sensitive data between multiple enclaves produced by the same development firm.

\smallskip
\noindent\hangindent 1em \textit{Attestation.} Attestation mechanism~\cite{mckeen2013innovative} is used to prove to a validator that an enclave has been correctly instantiated and when in that condition can then proceed to further establish a secure, authenticated connection for the data transmission. SGX provides two types of attestation: \textit{local attestation} and \textit{remote attestation}. In the former attestation, SGX facilitates the instructions to help an enclave to attest to another enclave on the same platform. In the latter one, SGX enables an enclave to prove a correct loading of code and data to another enclave that resides in a remote platform.

\begin{figure*}[htb!]
    \centering
    \includegraphics[width=0.98\linewidth]{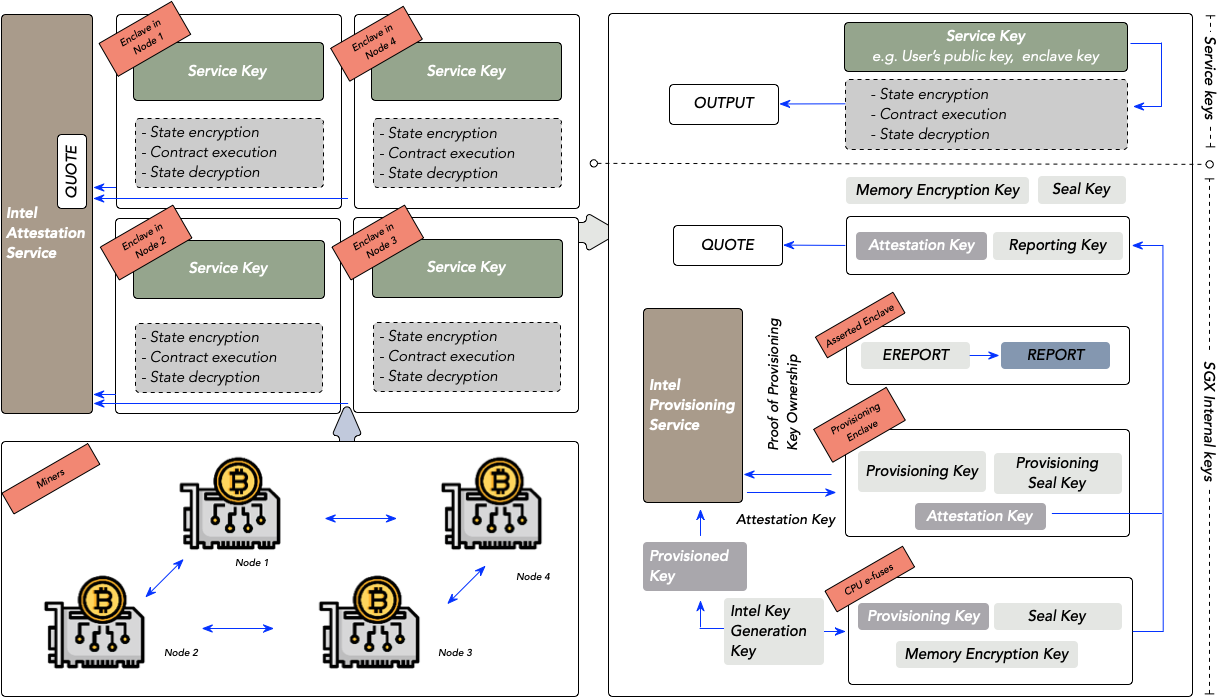}
   \caption{Key Usage in TEE-assisted Confidential Blockchain System.} 
    \label{fig:key-use}
\end{figure*}

\section*{Appendix D. A TCSC-based Voting Protocol}
\label{app-D}
In this part, we provide a detailed description of TCSC-based voting system that utilizes the Intex SGX. The protocol mainly consists of two sub-procedures: \textit{deployment stage} and \textit{execution stage}. We give details as follows.

\smallskip
\noindent\textbf{Deployment Stage.} In the deployment stage, all the operational code and the initial state are coded into a TCSC. This stage includes two steps.

\smallskip
\noindent\hangindent 1em\noindent\textit{Compile.} Firstly, contract binary codes are compiled into enclave codes. Since an enclave has only a small quantity of trusted zones for application code and data (the protected memory is 128MB, and only 96MB is usable for an enclave in the current version of Intel SGX~\cite{intel2016}), a contract has to determine the boundary of these zones and identify corresponding zones used for privacy-critical functionalities. In particular, the e-voting contract needs to define: \textit{the scope of secret states}, \textit{the scope of public states}, \textit{the approach to access secret states} and \textit{the approach to access external states}. Enclave Definition Language (EDL)~\cite{costan2016intel} defines trusted components, untrusted components, and corresponding interfaces between them, which takes charge of translation from contract code to enclave code. It provides two functionalities: Enclave Calls (ECALLs) and Outside Calls (OCALLs). ECALLs define the functions inside the enclave that are used to expose APIs for untrusted applications to call in. OCALLs specify untrusted functions outside the enclave where the enclave code is able to invoke. In our example, the total number of votes cast for a candidate cannot be revealed until the voting has ended. Thus, the total number of votes cast is defined at the access point ECALLs, and is thereby hidden from the public, and can only be revealed once the voting procedure has been completed. 

\smallskip
\noindent\hangindent 1em\textit{Load.} Afterwards, EDL files will load into an enclave, which is stored in the Enclave Page Cache (EPC). From a micro perspective, the first step is to call the $\mathsf{ECREATE}$ instruction for creating an enclave. This will allocate memory inside the Enclave Page Cache (EPC). Then, enclave code and data are added to pages in EPC by calling the $\mathsf{EADD}$ instruction. Finally, when the instruction $\mathsf{EINIT}$ completes successfully, an enclave’s $INIT$ attributes become true, and the above instructions cannot be used any more. After a successful deployment, the initial state and operational code of this contract will be replicated among blockchain nodes. This means the e-voting logic cannot be changed. But, the state of functionalities can be transferred to parties who have been granted permission with a message-call~\cite{wood2014ethereum}. 

\smallskip
\noindent\textbf{Execution Stage.} In the execution stage, voters call the deployed TCSC to finish the voting. Firstly, an enclave needs to fetch the current contract state from the blockchain. Then, the CPU executes the plaintext contract in the enclave mode. External attackers cannot obtain the knowledge of sensitive information since the Memory Encryption Engine (MEE) key never leaves TCB. A critical aspect of Intel SGX's functionality is that the code inside an enclave can access the particular enclave state by performing additional checks on memory semantics. Back to our example, confidential state (the encrypted number of votes cast for a candidate) will return only when the following four requirements are fulfilled:
\begin{itemize}
    \item[-] The processor runs in enclave mode;
    \item[-] The requested page is part of the same enclave;
    \item[-] The page access is through a correct virtual address;
    \item[-] The code semantics successfully pass the check.
\end{itemize}

In a word, the CPU is acting as a doorman in the TCSC, providing a hardware-based access control mechanism. After obtaining results from TEEs, the consensus algorithm starts to reach an agreement. To be specific, when a miner receives a newly mined block, he will re-execute all transactions inside the block to obtain the newly transferred state. Once enough blockchain miners receive the block and re-execute transactions, the voting results and the transactions triggering the contract execution will eventually reach the final agreement. When all the voting procedures have ended, the teller can fetch the final encrypted state and obtain the final voting result. In the meanwhile, the transactions can be used as evidence to trace the voter's behavior.


\end{document}